# Extremely high conductivity observed in the triple point topological metal MoP

Nitesh Kumar[1], Yan Sun[1], Michael Nicklas[1], Sarah J. Watzman[1,2], Olga Young[3], Inge Leermakers[3], Jacob Hornung[4,5], Johannes Klotz[4,5], Johannes Gooth[1], Kaustuv Manna[1], Vicky Süß[1], Satya N. Guin[1], Tobias Förster[4], Marcus Schmidt[1], Lukas Muechler[1,6], Binghai Yan[7], Peter Werner[8], Walter Schnelle[1], Uli Zeitler[3], Jochen Wosnitza[4,5], Stuart S. P. Parkin[8], Claudia Felser[1], Chandra Shekhar[1†]

[1]*Max Planck Institute for Chemical Physics of Solids, 01187 Dresden, Germany.*
[2]*Department of Mechanical and Aerospace Engineering, The Ohio State University, Columbus, Ohio 43210, USA.*
[3]*High Field Magnet Laboratory (HFML-EMFL) and Institute for Molecules & Materials, Radboud University, Toernooiveld 7, 6525 ED Nijmegen, The Netherlands.*
[4]*Dresden High Magnetic Field Laboratory (HLD-EMFL), Helmholtz-Zentrum Dresden-Rossendorf, 01328 Dresden, Germany.*
[5]*Institute for Solid-State and Material Physics, Technical University Dresden, 01062 Dresden, Germany.*
[6]*Department of Chemistry, Princeton University, Princeton, New Jersey 08544, USA.*
[7]*Department of Condensed Matter Physics, Weizmann Institute of Science, Rehovot 7610001, Israel.*
[8]*Max Planck Institute of Microstructure Physics, 06120 Halle, Germany.*

**Weyl and Dirac fermions have created much attention in condensed matter physics and materials science. Recently, several additional distinct types of fermions have been predicted. Here, we report ultra-high electrical conductivity in MoP at low temperature, which has recently been established as a triple point Fermion material. Here we show that the electrical resistivity is 6 nΩ cm at 2 K with a large mean free path of 11 microns. de Haas-van Alphen oscillations reveal spin splitting of the Fermi surfaces. In contrast to noble metals with similar conductivity and number of carriers, the magnetoresistance in MoP does not saturate up to 9 T at 2 K. Interestingly, the momentum relaxing time of the electrons is found to be more than 15 times larger than the quantum coherence time. This difference between the scattering scales shows that momentum conserving scattering dominates in MoP at low temperatures.**

Materials are conventionally divided into metals, semiconductors and insulators. Recently, through the lens of topology, materials can be reclassified as either topologically trivial or topologically non-trivial. Topologically non-trivial semimetals and metals exhibit novel, low-energy fermionic excitations of which the most extensively studied are the Dirac[1-4] and Weyl fermions[5-7]. Depending on the inherent symmetry of a particular compound, the crossing points can be several-fold degenerate[8]: two- and four-fold degenerate points are classified as Weyl and Dirac types, respectively. New materials going beyond Weyl and Dirac semimetals with higher band degeneracies have been proposed[8]. Dirac and Weyl semimetals in particular exhibit unusual transport properties, including high mobility[9-11], large magnetoresistance[9-11], anomalous Hall effect[12-14], and a chiral anomaly[14-16]. Although transport phenomena of Weyl and Dirac semimetals have been a topic of intense research, studies on metals with higher degenerate points, for instance, triple point topological metals, are still scarce.

In this article, we have explored the transport properties of the triple point metal MoP which exhibits an extremely low residual resistivity of 6 nΩcm and very long mean free path of 11 microns despite straightforward synthesis using conventional chemical vapor transport methods. The value of the residual resistivity is on a par with best metals known in the literature. Resistivity is a measure of loss of the charge carrier momentum by large-angle scattering with phonons and impurities. Therefore, large momentum relaxation reflects high resistivity and vice versa. On the other hand, electron-electron and small-angle scattering with phonons conserve momentum within the electron system and, therefore, hardly affect the resistivity[17,18]. The interplay among these scattering processes is at the heart of the hydrodynamic electron flow in materials such as graphene, $PdCoO_2$ and $WP_2$.[19-22] We show that these scattering processes play a crucial role in determining the excellent transport characteristics of the triple point Fermion metal MoP.

## Results

**Crystal structure and electronic structure.** MoP is a simple binary compound which was first predicted[23] and then confirmed experimentally to host topologically protected triple point (3-fold degenerate) fermions. It possesses a complex Fermi surface (FS)[24]. Besides the triple point fermions, pairs of Weyl nodes have also been observed experimentally[24]. However, the triple points and Weyl nodes are located far below $E_F$ (Ref [23,24] and Supplementary Fig. 1). MoP has a WC-type hexagonal crystal structure (Fig.1a) belonging to space group $P\bar{6}2m$ (No. 187) with lattice parameters $a = b = 3.22$ Å and $c = 3.19$ Å. Both Mo and P share the same coordination number and coordination environment of six in a trigonal prism. The Mo-P distances are very short and, consequently, the orbitals strongly hybridize with each other and give rise to a complex electronic structure with large bandwidths, as shown in Fig.1b. Near $E_F$, the bands are predominantly formed from the *d* orbitals of Mo with minor contributions from the *p* orbitals of P (see Supplementary Fig. 4).

The electrical resistivity, $\rho_{xx}$, of MoP crystals, that are grown via chemical vapor transport reactions using standard purity materials (99.999% for P and 99.95% for Mo), reaches an ultra-low value of 6 nΩ cm at $T = 2$ K, which is more than two times lower than Cu of similar purity[25] (Fig. 1d). The as-grown MoP crystals were analyzed using scanning transmission electron microscopy (high-angle annular dark-field HAADF-STEM). For these STEM studies, ~100 nm thick platelets were prepared by focused ion beam (FIB) milling, with an area of ~ 5 × ~ 5 μm$^2$. A representative high-resolution micrograph is shown in Fig. 1c, where Mo (bright large-dots) and P (small-dots) atoms are clearly visible. No evidence for defects is found.

The morphology of the crystals was analyzed by Laue x-ray diffraction (see Supplementary Fig. 5). The single crystals were then cut into bars whose faces were oriented such that $[2\bar{1}\bar{1}0]$ is along $\hat{\mathbf{x}}$, $[01\bar{1}0]$ is along $\hat{\mathbf{y}}$, and $[0001]$ is along $\hat{\mathbf{z}}$, as shown in the inset of Fig. 2d. Measurements of the resistivity were made with currents applied along each of these axes for several different crystals (Supplementary Table 1).

**Electrical and thermal transport measurements.** The temperature dependence of $\rho_{xx}$ is shown in Fig. 2a. In zero magnetic field, $\mathbf{B} = 0$ ($\mathbf{B} = \mu_0\mathbf{H}$), no significant difference in $\rho(T)$ was found for different axes or crystals (Supplementary Fig. 6), notwithstanding the hexagonal symmetry. The material exhibits very low resistivity values at low temperature, which is independent of the crystal orientation. As an example, we consider crystal S3 (Fig. 2a, $\mathbf{I} \parallel \hat{\mathbf{x}}$,). The measured values of $\rho_{xx}$ are 6 nΩ cm at 2 K and 8.2 μΩ cm at 300 K, thereby resulting in a very large residual resistivity ratio, RRR = $\rho_{xx}$ (300 K) /$\rho_{xx}$ (2 K) = 1370. RRR values for other crystals are given in Supplementary Table 1. The RRR is higher than found, for example, for WC, a TPF semimetal ($RRR$ ~ 85)[26] and for the TaAs family of Weyl semimetals (RRR ~ 115)[11,27], but lower than, for example, high-quality samples of the Dirac semimetal $Cd_3As_2$ (RRR = 4100)[10]. On the other hand, the residual resistivity, $\rho_0$ of each of these compounds is much higher than that of MoP (~ 0.35 μΩcm for WC (Ref.[26]), ~ 0.63 μΩcm for NbP (Ref.[11]) and ~ 21 nΩcm for $Cd_3As_2$ (Ref.[10])). The temperature dependence of $\rho_{xx}$ in $\mathbf{B} = 0$ and 14 T (along $\hat{\mathbf{z}}$) is compared in Fig 2b showing that a significant magnetoresistance sets (for other crystals, see supplementary Fig. 9) in below 40 K. This behavior is quite different from conventional metals, such as copper. We fit the low temperature regime of the zero field resistivity according to $\rho_{xx} = \rho_0 + AT^2 + BT^5$, where $\rho_0$ ~ 6.0 ×$10^{-9}$ Ωcm, $A$ ~ 2.3×$10^{-12}$ ΩcmK$^{-2}$ and $\mathbf{B}$ ~3.6×$10^{-16}$ ΩcmK$^{-5}$, respectively. The quadratic term is a ubiquitous feature of Fermi liquids, originating from slow momentum decay via electron-electron scattering. The $T^5$-term is usually associated with electron phonon scattering. The value of $A$ in MoP exceeds that of $A_{ab}$ in $PdCoO_2$, a compound which has shown signatures of a hydrodynamic electron fluid. Importantly, for the free electron gas like aluminum the value of $A$ is one order smaller[28].

**Kadowaki-Woods ratio and Wiedemann-Franz law.** To gain further insight into the correlation between charge carriers at low temperatures in MoP, its specific heat $C_p$ was measured as a function of temperature (Supplementary Fig. 7). The Sommerfeld coefficient, $\gamma$, and Debye temperature, $\Theta_D$ are 2.37 mJ mol$^{-1}$ K$^{-2}$ and 573 K, respectively. In correlated metals, the pre-factor $A$ of the $T^2$-dependent electrical resistivity and $\gamma$ are fundamentally linked by the modified Kadowaki-Woods ratio (KWR)[29], because electron-electron interactions enhance both coefficients:

$$\frac{Af_{dx}}{\gamma^2} = \frac{81}{4\pi\hbar k_B^2 e^2} = 1.25 \times 10^{118}\ \Omega \mathrm{mkg^{-4}m^{-9}s^6K^2}, \qquad (1)$$

where $f_{dx}$ is a materials specific quantity, $\hbar$ denotes the reduced Planck constant, $k_B$is the Boltzmann constant, and e is the elementary charge. We calculate $f_{dx} = \sqrt[3]{3n^7/\pi^4\hbar^6}$ ,[29]. This gives a KWR of $5.1 \times 10^{118}$ $\Omega$mkg$^{-4}$m$^{-9}$s$^6$K$^2$, which is comparable to that calculated for correlated electron systems[29]. In fact, signatures for electron-electron interaction have previously been found in many semimetals such as in Sb[30], Bi[31], $WP_2$[22] and $PdCoO_2$[21] at low temperatures. The KWR of MoP highlights once more the importance of the electron-electron interaction in semimetals at low temperatures.

To address the interactions and accompanied scattering mechanism in MoP in more detail, we have next measured the thermal conductivity $\kappa$, which is related to the electrical conductivity, $\sigma$ through the Wiedemann-Franz (WF) law. According to the WF law $\kappa_{el} = L_0\sigma T$, where $\kappa_{el}$ is the electronic thermal conductivity and $L_0$ is a universal constant known as the Sommerfeld value of the Lorenz number. The measured temperature dependence of the total thermal conductivity $\kappa(T)$ of MoP (crystal S3 with a temperature gradient along $\hat{\mathbf{x}}$.) increases on decreasing temperature and reaches a value of 3000 WK$^{-1}$m$^{-1}$ at 23 K as shown in Supplementary Fig. 7c. Using the Debye model (see Supplementary Note 6 for details), we extracted $\kappa_{el}$ and calculated the normalized Lorenz number ($L/L_0$) as a function of temperature as shown in Fig. 2c. While the WF law is obeyed at high temperatures, below 40 K, we observe a sudden decrease of $L/L_0$. At 23 K, the value of $L/L_0$ is ~ 0.4, corresponding to a decrease by 60 %. Such a large deviation from the WF law can also be seen in many conventional metals [32]. The main reason for the deviation from the WF law is the presence of inelastic scattering at intermediate temperatures which degrade the electrical and thermal currents differently. We will discuss various scattering mechanisms in MoP in the following sections. Interestingly, such a large violation of the WF law was also recently observed in $WP_2$ and $PtSn_4$; both of them showing evidence towards hydrodynamic fluid [22,33]. However, the correlation between these two observations is yet to be established experimentally.

**Fermi surface topology.** To get more quantitative information about the scattering times, we measured the Hall resistivity, $\rho_{yx}$, at various temperatures (Supplementary Fig. 8), which can be described by the charge-carrier transport from a single hole band over the entire temperature range from 2 to 300 K. The mobility, $\mu$, and carrier density, $n$, obtained from $\rho_{yx}$ are $1.4 \times 10^4$ cm$^2$/Vs, and $3.9 \times 10^{22}$ cm$^{-3}$, respectively, at 2 K, as shown in Fig. 2d and Supplementary Fig. 8d for crystal V2. A small increase in the charge-carrier density with increasing temperature, from $3.9 \times 10^{22}$ cm$^{-3}$ at 2 K, to $1.1 \times 10^{23}$ cm$^{-3}$ at 300 K, is found. However, a much larger decrease in the mobility of ~1,000 times is observed over the same temperature range. We studied various MoP crystals grown in two batches (S and V) which are listed in Supplementary Table 1. We find that all the crystals possess a high mobility and large conductivity, $\sigma_{xx}$, as shown in Supplementary Fig. 6b. The values of $\sigma_{xx}$ are mainly controlled by the mobility which is directly proportional to the momentum-relaxing scattering time, $\tau_{mr}$, and momentum relaxing length, $l_c$, which we determined to be $1.2 \times 10^{-11}$ s and 11 μm at 2 K, respectively, for the crystal V1 (see below for more details).

$\tau_{mr}$ is also called the Drude scattering time, $\tau_{mr} = \mu\, m^*/e$. The momentum of the charge carriers is therefore conserved across at least $10^4$ unit cells of MoP.

To get an estimate of the total amount of scattering events beyond momentum relaxing processes in MoP at low $T$, we have measured de-Haas van-Alphen (dHvA) oscillations via magnetic torque. This allows for the extraction of the quantum coherence time, which is extremely sensitive to any scattering events that thermalize the electron system. To evaluate these experiments, we have first constructed the FSs with the help of *ab-initio* electronic structure calculations[34,35], which are shown in Fig. 3. There are three kinds of FSs; tiny droplet-type electron pockets elongated along the $\Gamma$-$A$ direction (Fig. 3a), disc-shaped hole pockets located around the center of the Brillouin zone (BZ) (Fig. 3b), and typical metallic open FSs extending over almost the whole BZ (Fig. 3c). These FSs are consistent with earlier ARPES experiments[24]. Due to the non-centrosymmetric structure of MoP, all the bands are spin-split[36,37] resulting in pairs of FSs with similar shapes but different sizes. This spin splitting in a non-magnetic compound (with time reversal symmetry) such as MoP could play a crucial role in the transport properties since it limits back-scattering. We estimate from the band structure calculations that the charge carrier concentration is ~$2.8 \times 10^{22}$ cm$^{-3}$, which is reasonably close to the measured values ($2.9$-$4.3 \times 10^{22}$ cm$^{-3}$ from Supplementary Table 1). These carriers arise largely from the metallic open FSs, with merely ~ 4 % from the closed hole pockets, and a negligible contribution from the electron pockets (< 1 %). The Fermi velocity, $\mathbf{v}_F$, and the orbital distribution of different FSs are shown in Supplementary Fig. 3 which reveal that the charges in the open FSs and electron pockets have similar $\mathbf{v}_F$. Having established the Fermiology of MoP in theory, torque measurements for $\mathbf{B} \parallel \hat{\mathbf{x}}$ at various temperatures and the angular dependence in the $xz$ plane at 1.2 K are shown in Fig. 4a and Supplementary Figs 10-12. The dHvA oscillations, which are readily visible up to 20 K are periodic in $1/B$ (Fig. 4b). The frequencies, $F$, of these oscillations are directly related to the extremal cross-sectional areas $A_F$ of the FS perpendicular to **B** via the Onsager relation $F = (\Phi_0/2\pi^2)A_F$, where $\Phi_0 = h/2e$ ($= 2.068\times10^{-15}$ Wb) is the magnetic flux quantum and $h$ is the Planck constant. Fourier transforms (Fig. 4c, e) of the torque data show fundamental frequencies at $F_{\alpha 3} = 855$ T, $F_{\varepsilon 1} = 2120$ T, $F_\eta = 3550$ T and $F_\xi = 14560$ T and such large frequency is observed only in some selected materials [21,38,39]. This reveals large Fermi pockets usually not expected in typical Dirac[10] and Weyl semimetals[9,11]. Our observations are in very good agreement with the frequencies derived from our band structure calculations. The dependence of the dHvA frequencies on the orientation ($\theta$) of **B** in the xz-plane at 1.2 K is shown in Fig. 4f and Supplementary Fig. 11 and displays a $1/\cos\theta$ behavior at low angles that is expected for a predominantly 2D FS. Since $F_\eta$ corresponds to the open FSs, these oscillations can only be observed when **B** is close to $\hat{\mathbf{z}}$ ($\theta = 90^o$) and, for this reason, we measured the temperature dependence torque signal at $\theta = 75^o$ (Supplementary Fig. 12). Noticeably, the observed oscillations exhibit a clear beating pattern (Fig 4b) due to the very close frequencies of $\alpha_2$ (810 T) & $\alpha_3$ (855 T). These frequencies correspond to spin-split bands due to the lack of inversion symmetry[36,37]. Another spin-split pair of frequencies are $\varepsilon_1$ (2120 T) and $\varepsilon_2$ (2220 T). Effective mass, $m^*$, $A_F$, Fermi wave vector, $k_F$, and Fermi velocity, $\mathbf{v}_F$, are calculated from the temperature dependence of the oscillation amplitudes (Fig. 4d) using the Lifshitz–Kosevich formula $R_T = X/\sinh(X)$, where $X = 14.69m^*T/B$ and $B$ is the average field, and from the Onsager relation, $k_F = \sqrt{A_F/\pi}$ , $v_F = k_F h/2\pi m^*$, respectively[40]. The resulting values of $m^*$, $A_F$, $k_F$, $v_F$ corresponding to the different FS pockets are summarized in Supplementary Table 2. Among the pockets, these values are highest for the open FSs for which we find: $m^* = 1.12m_0$, $A_F = 1.39$ Å$^{-2}$, $k_F = 0.67$ Å$^{-1}$, $v_F = 6.9\times 10^5$ ms$^{-1}$. The experimental values are consistent with those calculated theoretically. $m^*$ for the different bands varies strongly (from $0.23m_0$ to $1.12m_0$, see Supplementary Table 2) which underlines the complex band structure of MoP. All the bands are spin non-degenerate and the Fermi wave vectors of the spin-split bands $\alpha$ and $\varepsilon$ are separated by 0.004 Å$^{-1}$ ($k_{F\alpha3}$ - $k_{F\alpha2}$ = (0.161-0.157) Å$^{-1}$) and 0.006 Å$^{-1}$ ($k_{F\varepsilon2}$ - $k_{F\varepsilon1}$ = (0.260-0.254) Å$^{-1}$), respectively.

**Momentum relaxing and momentum conserving scattering times.** The quantum scattering time is then given by$\tau_D = \hbar/(2\pi k_B T_D)$ or $1.22 \times 10^{-12}/T_D$, which we estimate as $\tau_D = 7.6 \times 10^{-14}$ s at 2 K from the Dingle temperature, $T_D \sim (1.6 \pm 0.15)$ K. The value of $T_D$ was estimated from the linear slope of the Dingle plot as shown in Supplementary Fig. 13. The oscillatory behavior of points around the fitted straight line is due to the presence of beating patterns, which are an intrinsic property of the material. At first approximation, the momentum conserving scattering time $\tau_{mc}$ can be estimated by applying Mathiessen's rule $1/\tau_D = 1/\tau_{mr} + 1/\tau_{mc}$. Consequently, $\tau_{mc}$ is found to be same order of magnitude as $\tau_D$ since $1/\tau_{mr} \ll 1/\tau_{mc}$ , showing that most of the scattering in MoP at low temperatures conserves the total momentum of the electrons. The value of $\tau_{mr}/\tau_{mc}$ is found to be 15 for MoP.

## Discussion

For hydrodynamic charge flow, the ratio of these scattering times, $\tau_{mr}/\tau_{mc} \gg 1$ is the key feature[41]. Similar to the known systems with hydrodynamic charge flow in $PdCoO_2$ [21] and $WP_2$[22], the ratio of $\tau_{mr}/\tau_{mc}$ MoP is very high. Interestingly, these compounds have different dimensionalities but they share common properties such as high conductivity, high magnetoresistance, high mean free path and electron-electron correlations. MoP, therefore, is a strong candidate to meet the requirements for three-dimensional hydrodynamic charge flow. However, despite the similarities in transport characteristics, an important question remains open and requires careful theoretical and experimental investigation in the future: Which microscopic processes make the momentum currents in these materials long lived at low temperatures, even in the limit where the electrical resistivity is limited by impurity scattering?

In summary, MoP is a highly unusual topological metal with high conductivity and mobility, despite no elaborate efforts were taken to eliminate any type of disorder and impurities. The open Fermi surface covers a significant fraction of the volume of the first Brillouin zone and, therefore, is responsible for the high conductivity in MoP. At low temperatures, the compound exhibits an unusually high suppression of momentum relaxation, which is the manifestation of timescales of momentum relaxing and momentum conserving processes being different by two orders of magnitude. MoP is one member of the conventional class of materials that hosts unconventional fermions providing an exciting platform for hydrodynamic electron flow in solids.

## Methods

**Single crystal growth and characterizations.** The single crystals of MoP were grown via a simple chemical vapor transport method using iodine as a transport agent. First, MoP polycrystalline powder was synthesized by a direct reaction of molybdenum (Alfa-Aesar, 99.95%) and red phosphorus (Alfa-Aesar, 99.999%) sealed in an evacuated fused silica tube. The sealed tube was heated to 873 K and then to 1073 K. Starting from this powder with iodine, the single crystals were grown in a two-zone furnace at a temperature of 1273 K ($T_2$) and 1173 K ($T_1$). After several weeks, the ampoule was removed and then quenched in water. Plate-like crystals, of size 0.5-1 mm, were obtained and were further characterized by x-rays and energy dispersive spectroscopy (EDS). The orientation and crystal structure were investigated by Laue x-ray diffraction.

**Electrical and thermal transport measurements.** The transport measurements of several MoP single crystals were performed in a physical property measurement systems (PPMS, Quantum Design, ACT option, specific heat option, external low resistivity AC bridge set-up and customized external thermal transport set-up[42]).

**Magnetic torque measurement.** The magnetic torque measurements were performed in an 18 T superconducting magnet at the Dresden High Magnetic Field Laboratory (HLD-EMFL), Helmholtz-Zentrum Dresden-Rossendorf, Germany and up to 35 T in static magnetic fields at the High Field Magnet Laboratory HFML-RU/FOM in Nijmegen, The Netherlands. We paid extra attention to estimate Dingle temperature from quantum oscillations because oscillation amplitude is largely smeared by magnetic field inhomogeneity and limited sampling rate (see Supplementary Fig. 14 and Supplementary Note 8). Besides these, an overlapping field windows with constant width in $1/\mu_0 H$ for FFT amplitude was selected to minimize the beating effect in Dingle plot. A more detailed discussion can be found in supplementary information.

**Band structure calculations.** The electronic structure was calculated by first principles calculations based on density functional theory, using the projected augmented wave method as implemented in the Vienna ab-initio Simulation Package (VASP)[35]. The exchange and correlation energy was considered in the generalized gradient approximation (GGA) with the Perdew-Burke-Ernzerhof-based (PBE) density functional[43]. The tight binding model Hamiltonian was calculated by projected Bloch states onto maximally localized Wannier functions (MLWFs)[44].

**Data availability**. The data that support the findings of this study are available from the corresponding author C.S. upon request.

**References:**

1 Wang, Z. *et al.* Dirac semimetal and topological phase transitions in $A_3$Bi ($A$=Na, K, Rb). *Phys. Rev. B* **85**, 195320 (2012).

2 Neupane, M. *et al.* Observation of a three-dimensional topological Dirac semimetal phase in high-mobility $Cd_3As_2$. *Nat. Commun.* **5**, 3786 (2014).

3 Liu, Z. K. *et al.* A stable three-dimensional topological Dirac semimetal $Cd_3As_2$. *Nat. Mater.* **13**, 677-681 (2014).

4 Liu, Z. K. *et al.* Discovery of a Three-Dimensional Topological Dirac Semimetal, $Na_3Bi$. *Science* **343**, 864-867 (2014).

5 Weng, H., Fang, C., Fang, Z., Bernevig, B. A. & Dai, X. Weyl Semimetal Phase in Noncentrosymmetric Transition-Metal Monophosphides. *Phys. Rev. X* **5**, 011029 (2015).

6 Xu, S.-Y. *et al.* Discovery of a Weyl fermion state with Fermi arcs in niobium arsenide. *Nat. Phys.* **11**, 748-754 (2015).

7 Liu, Z. K. *et al.* Evolution of the Fermi surface of Weyl semimetals in the transition metal pnictide family. *Nat. Mater.* **15**, 27-31 (2016).

8 Bradlyn, B. *et al.* Beyond Dirac and Weyl fermions: Unconventional quasiparticles in conventional crystals. *Science* **353**, aaf5037 (2016).11

9 Ali, M. N. *et al.* Large, non-saturating magnetoresistance in $WTe_2$. *Nature* **514**, 205-208 (2014).

10 Liang, T. *et al.* Ultrahigh mobility and giant magnetoresistance in the Dirac semimetal $Cd_3As_2$. *Nat. Mater.* **14**, 280-284 (2015).

11 Shekhar, C. *et al.* Extremely large magnetoresistance and ultrahigh mobility in the topological Weyl semimetal candidate NbP. *Nat. Phys.* **11**, 645-649 (2015).

12 Burkov, A. A. Anomalous Hall Effect in Weyl Metals. *Phys. Rev. Lett.* **113**, 187202 (2014).

13 Suzuki, T. *et al.* Large anomalous Hall effect in a half-Heusler antiferromagnet. *Nat. Phys.* **12**, 1119-1123 (2016).

14 Shekhar, C. *et al.* Anomalous Hall effect in Weyl semimetal half-Heusler compounds RPtBi (R = Gd and Nd). *Proc. Natl. Acad. Sci. U.S.A.* **115**, 9140-9144 (2018).

15 Hirschberger, M. *et al.* The chiral anomaly and thermopower of Weyl fermions in the half-Heusler GdPtBi. *Nat. Mater.* **15**, 1161-1165 (2016).

16 Felser, C. & Yan, B. Weyl semimetals: Magnetically induced. *Nat. Mater.* **15**, 1149-1150 (2016).

17 Gurzhi, R. N., Kalinenko, A. N. & Kopeliovich, A. I. Hydrodynamic effects in the electrical conductivity of impure metals. *Sov. Phys. JETP* **69**, 863-870 (1989).

18 Steinberg, M. Viscosity of the electron gas in metals. *Phys. Rev.* **109**, 1486 (1958).

19 Bandurin, D. A. *et al.* Negative local resistance caused by viscous electron backflow in graphene. *Science* **351**, 1055 (2016).

20 Crossno, J. *et al.* Observation of the Dirac fluid and the breakdown of the Wiedemann-Franz law in graphene. *Science* **351**, 1058-1061 (2016).

21 Moll, P. J. W., Kushwaha, P., Nandi, N., Schmidt, B. & Mackenzie, A. P. Evidence for hydrodynamic electron flow in $PdCoO_2$. *Science* **351**, 1061-1064 (2016).

22 Gooth, J. *et al.* Thermal and electrical signatures of a hydrodynamic electron fluid in tungsten diphosphide. *Nat. Commun.* **9**, 4093 (2018).

23 Zhu, Z., Winkler, G. W., Wu, Q., Li, J. & Soluyanov, A. A. Triple Point Topological Metals. *Phys. Rev. X* **6**, 031003 (2016).

24 Lv, B. Q. *et al.* Observation of three-component fermions in the topological semimetal molybdenum phosphide. *Nature* **546**, 627-631 (2017).

25 Nakane, H., Watanabe, T., Nagata, C., Fujiwara, S. & Yoshizawa, S. Measuring the temperature dependence of resistivity of high purity copper using a solenoid coil (SRPM method). *IEEE Trans. Instrum. Meas.* **41**, 107-110 (1992).

26 He, J. B. *et al.* Magnetotransport properties of the triply degenerate node topological semimetal tungsten carbide. *Phys. Rev. B* **95**, 195165 (2017).

27 Arnold, F. *et al.* Negative magnetoresistance without well-defined chirality in the Weyl semimetal TaP. *Nat. Commun.* **7** (2016).

28 Buyco, E. H. & Davis, F. E. Specific heat of aluminum from zero to its melting temperature and beyond. Equation for representation of the specific heat of solids. *J. Chem. Eng. Data* **15**, 518-523 (1970).

29 Jacko, A. C., Fjaerestad, J. O. & Powell, B. J. A unified explanation of the Kadowaki-Woods ratio in strongly correlated metals. *Nat. Phys.* **5**, 422-425 (2009).

30 Chopra, V., Ray, R. K. & Bhagat, S. M. Low-temperature resistivity of Bi and its alloys. *Phys. Status Solidi A* **4**, 205-214 (1971).

31 Collan, H. K., Krusius, M. & Pickett, G. R. Specific Heat of Antimony and Bismuth between 0.03 and 0.8 K. *Phys. Rev. B* **1**, 2888-2895 (1970).

32 Berman, R. & MacDonald, D. K. C. The thermal and electrical conductivity of copper at low temperatures. *Proc. Roy. Soc. London Ser. A* **211**, 122 (1952).

33 Fu, C. *et al.* Thermoelectric signatures of the electron-phonon fluid in $PtSn_4$. *arXiv preprint arXiv:1802.09468* (2018).

34 Kresse, G. & Hafner, J. *Ab initio* molecular dynamics for open-shell transition metals. *Phys. Rev. B* **48**, 13115-13118 (1993).

35 Kresse, G. & Furthmüller, J. Efficiency of ab-initio total energy calculations for metals and semiconductors using a plane-wave basis set. *Comput. Mater. Sci.* **6**, 15-50 (1996).

36 Ishizaka, K. *et al.* Giant Rashba-type spin splitting in bulk BiTeI. *Nat. Mater.* **10**, 521-526 (2011).

37 Nitta, J., Akazaki, T., Takayanagi, H. & Enoki, T. Gate Control of Spin-Orbit Interaction in an Inverted $In_{0.53}Ga_{0.47}As/In_{0.52}Al_{0.48}As$ Heterostructure. *Phys. Rev. Lett.* **78**, 1335-1338 (1997).

38 Tan, B. S. *et al.* Unconventional Fermi surface in an insulating state. *Science* (2015).

39 Joseph, A. S., Thorsen, A. C. & Blum, F. A. Low-Field de Haas-van Alphen Effect in Gold. *Phys. Rev.* **140**, A2046-A2050 (1965).

40 Shoenberg, D. *Magnetic Oscillations in Metals*. (Cambridge Univ. Press, 1984).

41 Principi, A. & Vignale, G. Violation of the Wiedemann-Franz Law in Hydrodynamic Electron Liquids. *Phys. Rev. Lett.* **115**, 056603 (2015).

42 Stockert, U. *et al.* Thermopower and thermal conductivity in the Weyl semimetal NbP. *J. Phys. Condens. Matter* **29**, 325701 (2017).

43 Perdew, J. P., Burke, K. & Ernzerhof, M. Generalized Gradient Approximation Made Simple. *Phys. Rev. Lett.* **77**, 3865-3868 (1996).

44 Marzari, N. & Vanderbilt, D. Maximally localized generalized Wannier functions for composite energy bands. *Phys. Rev. B* **56**, 12847-12865 (1997).

**Acknowledgements**

We acknowledge Prof. Andrew P. Mackenzie for helpful discussion, Dr. Horst Blumtritt for preparing the samples for STEM investigations and Dr. Ulrike Stockert for help in thermal conductivity measurements. This work was financially supported by the ERC Advanced Grant No. 742068 'TOPMAT'. We also acknowledge the support of the High Field Magnet Laboratory Nijmegen (HFML-RU/FOM) and HLD at HZDR, members of the European Magnetic Field Laboratory (EMFL).

**Author contributions**

C.S. designed the experiments. N.K. and C.S. performed transport measurements with inputs from M.N. and W.S. V.S. and M.S. grew the single crystals. C.S. and K.M. performed Laue x-ray diffraction experiments. I.L., O.Y., U.Z., and T.F. performed high magnetic field torque and transport measurements. J.H., J.K., and J.W. designed and performed the torque measurements in the superconducting 18 T magnet. Y.S. with inputs from L.M. and B.Y. carried out theoretical calculations. J.G., S.W. and S.G. performed thermal measurements. P.W. performed microstructural characterizations C.S., N.K., S.S.P.P. and C.F wrote the manuscript with inputs from all authors. C.F. supervised the project.

**Competing interests**

The authors declare no competing interests.

**Correspondence**

Correspondence and requests for materials should be addressed to C.S. (email: shekhar@cpfs.mpg.de).

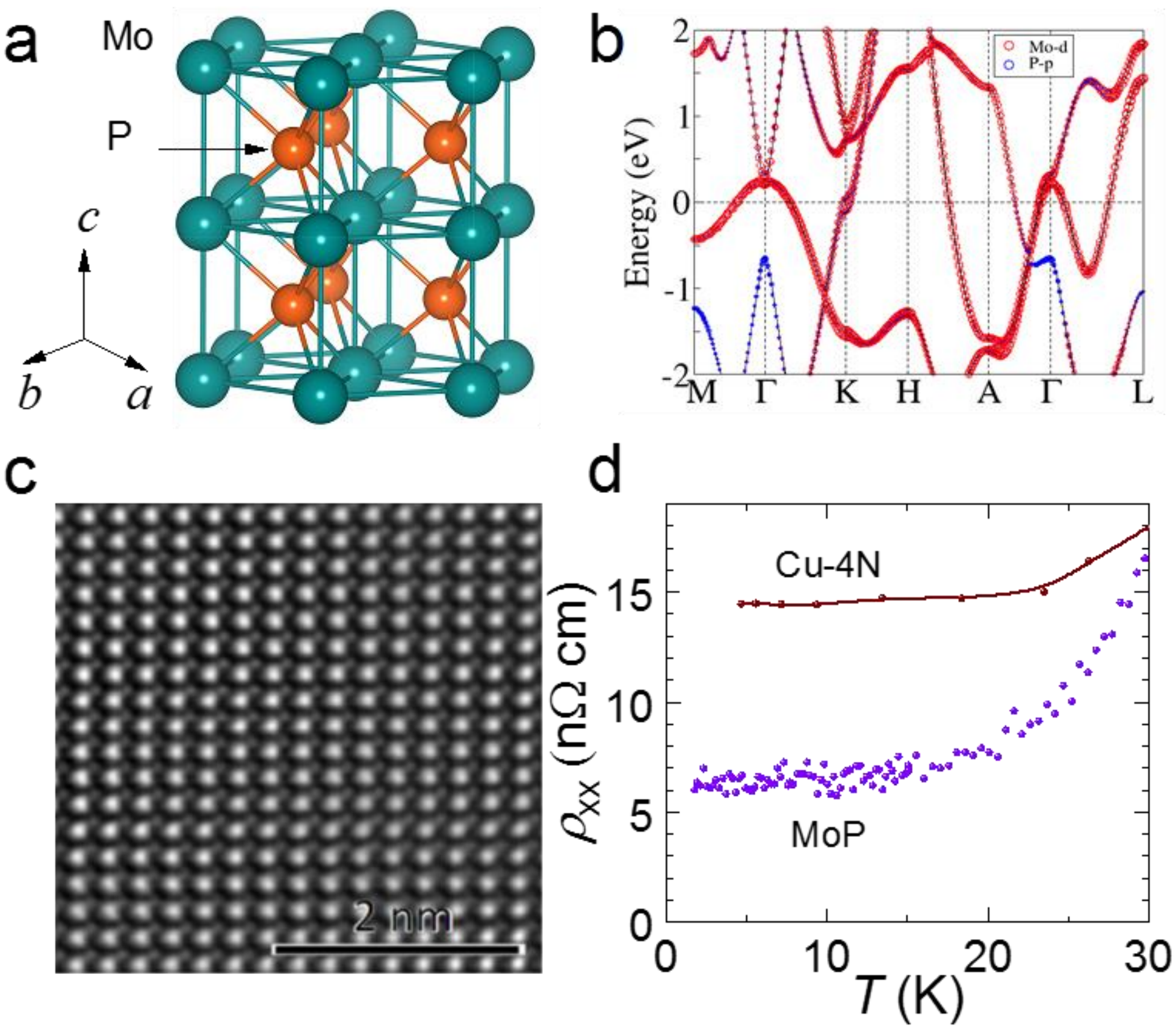

**Figure 1 | Crystal structure, electronic band structure, high-resolution scanning transmission electron microscopic (HR-STEM) image and resistivity of MoP.** (**a**) Schematic of the hexagonal crystal structure of MoP in which the Mo and P atoms are shown as green and orange spheres, respectively. (**b**) Energy dispersed band-structure along the high symmetry directions, including spin-orbit coupling (SOC). Mo *d*-orbitals contributions (red) dominate at $E_F$. (**c**) HR-STEM image along [110], showing Mo (bright large-dots) and P (small-dots). (**d**) Temperature dependent resistivity, $\rho_{xx}$, of MoP compared with 99.99 % (4N) pure Cu.

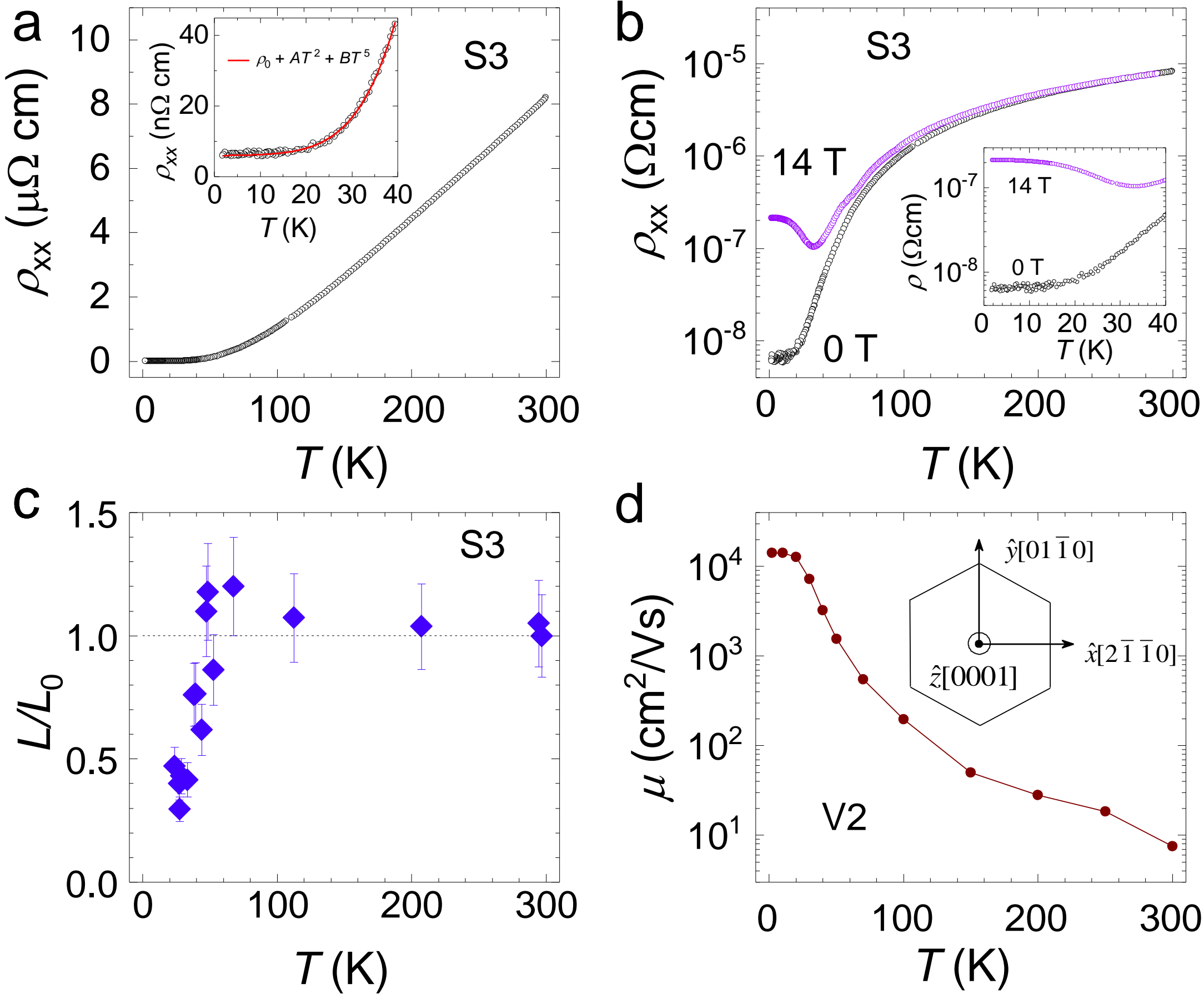

**Figure 2 | Resistivity, Lorenz number and mobility of MoP.** (**a**) Temperature dependent resistivity, $\rho_{xx}$ for $\mathbf{I}||\hat{\mathbf{x}}$. The inset shows the best fit with $\rho_{xx} = \rho_0 + AT^2 + BT^5$ for the region $T < 50$ K, where the values of $\rho_0$, $A$ and $B$ are $6.0 \times 10^{-9}$ Ωcm, $2.3 \times 10^{-12}$ ΩcmK$^{-2}$ and $3.6 \times 10^{-16}$ ΩcmK$^{-5}$, respectively. (**b**) Temperature dependent $\rho_{xx}$ at 0 and 14 T when $\mathbf{I}||\hat{\mathbf{x}}$ and $\mathbf{B}||\hat{\mathbf{z}}$. The inset shows a significant magnetoresistance below 40 K. (**c**) Temperature dependent normalized Lorenz number, $L/L_0$, where $L_0$ is the Lorenz number. The error bar comes from the error in the thermal conductivity measurement (See Supplementary Fig. 7). A strong violation of the Wiedemann-Franz law is observed below 40 K. (**d**) Temperature dependent charge-carrier mobility, $\mu$, of MoP. The inset of **d** shows the crystallographic directions in the hexagonal representation in which $\hat{\mathbf{x}}$, $\hat{\mathbf{y}}$ and $\hat{\mathbf{z}}$ are defined as $[2\bar{1}\bar{1}0]$, $[01\bar{1}0]$ and $[0001]$, respectively.

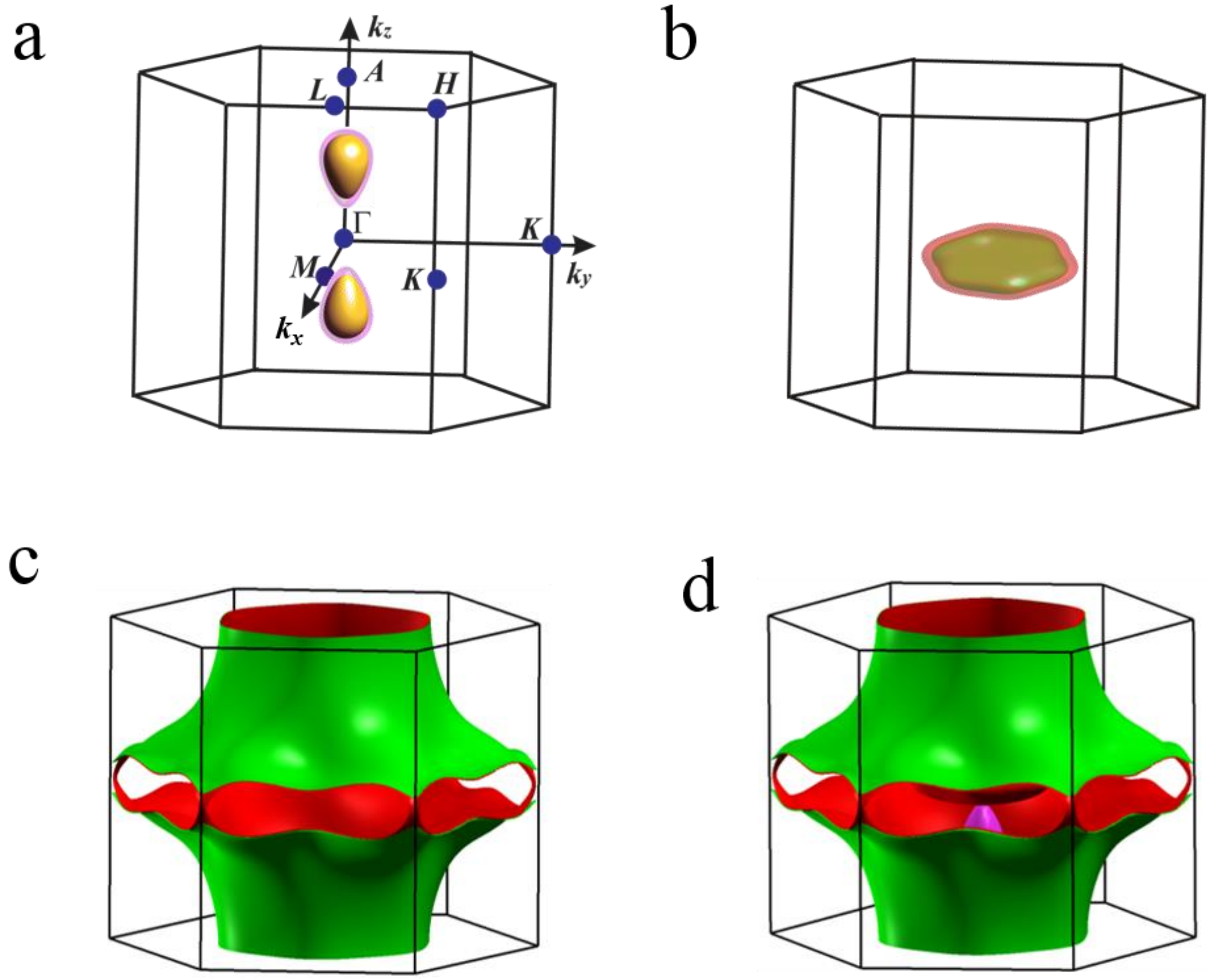

**Figure 3 | Shape of the Fermi surface (FS) in the first Brillouin zone (BZ) of MoP**. (**a**) Small droplet-type electron pockets which contribute negligibly to the charge-carrier density. (**b**) Flat hole pocket at the center of the BZ. (**c**) Open FSs that extend over the entire BZ. (**d**) Electron pockets, hole pocket, and open FSs combined together in the BZ. Due to the non-centrosymmetric structure of MoP, the pockets as well as the open FSs are spin split and appear in pairs.

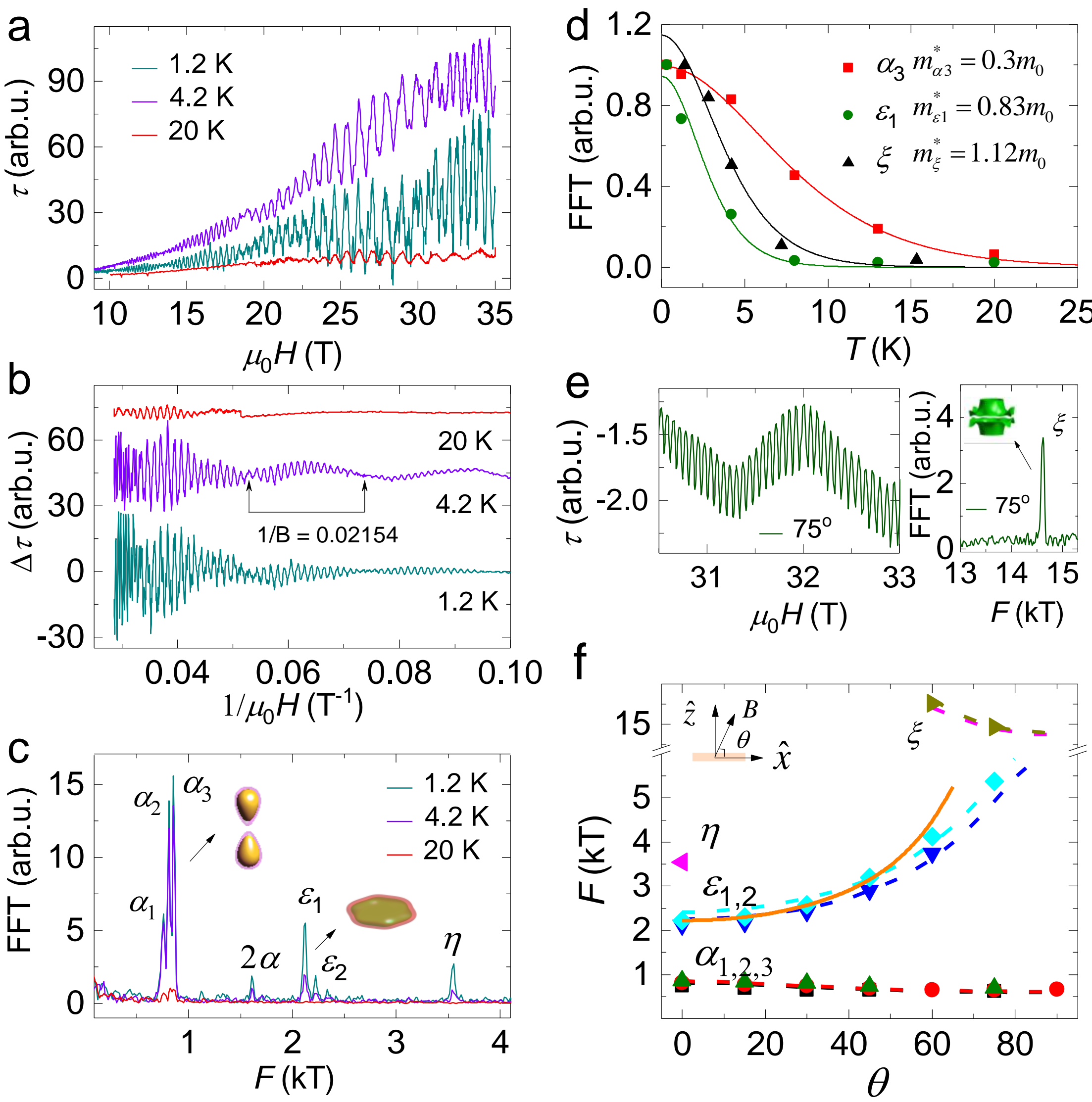

**Figure 4 | Quantum oscillations, fast Fourier transform (FFT), effective mass, and angular dependent oscillation frequencies of MoP for the crystal V1.** (**a**) de-Haas van-Alphen (dHvA) oscillations from magnetic torque measurements, when $\mathbf{B}||\hat{\mathbf{x}}$, at various temperatures. (**b**) Oscillatory components after subtracting a 3rd order polynomial from the data. Beating patterns (marked by arrows) are clearly visible that have a frequency of 45 T, i.e. the difference in frequencies between $\alpha_2$ and $\alpha_3$. (**c**) FFTs from 10 - 35 T showing different frequencies corresponding to the different pockets involved in the quantum oscillations. (**d**) Temperature dependence of dHvA amplitudes from FFTs and LK fits giving the effective masses corresponding to different frequencies, $\alpha_3$, $\varepsilon_1$ and $\eta$. (**e**) Magnetic torque signal ($\theta = 75^o$) showing very dense dHvA oscillations: these are visible only when **B** is nearly along $\hat{\mathbf{z}}$ (left panel). The corresponding FFT exhibits a single frequency of 14.6 kT (right panel). (**f**) Angular dependent frequencies when $B$ is rotated from $\hat{\mathbf{x}}$ ($\theta = 0^o$) to $\hat{\mathbf{z}}$ ($\theta = 90^o$) where the symbols and dotted lines are experimental and calculated values, respectively. The solid line (orange color) depicts a $1/\cos\theta$ behavior. The inset shows the rotation geometry of **B** with respect to the $\hat{\mathbf{x}}$ and $\hat{\mathbf{z}}$ directions.

## SUPPLEMENTARY INFORMATION

**Kumar *et al*. Extremely high conductivity observed in the triple point topological metal MoP**

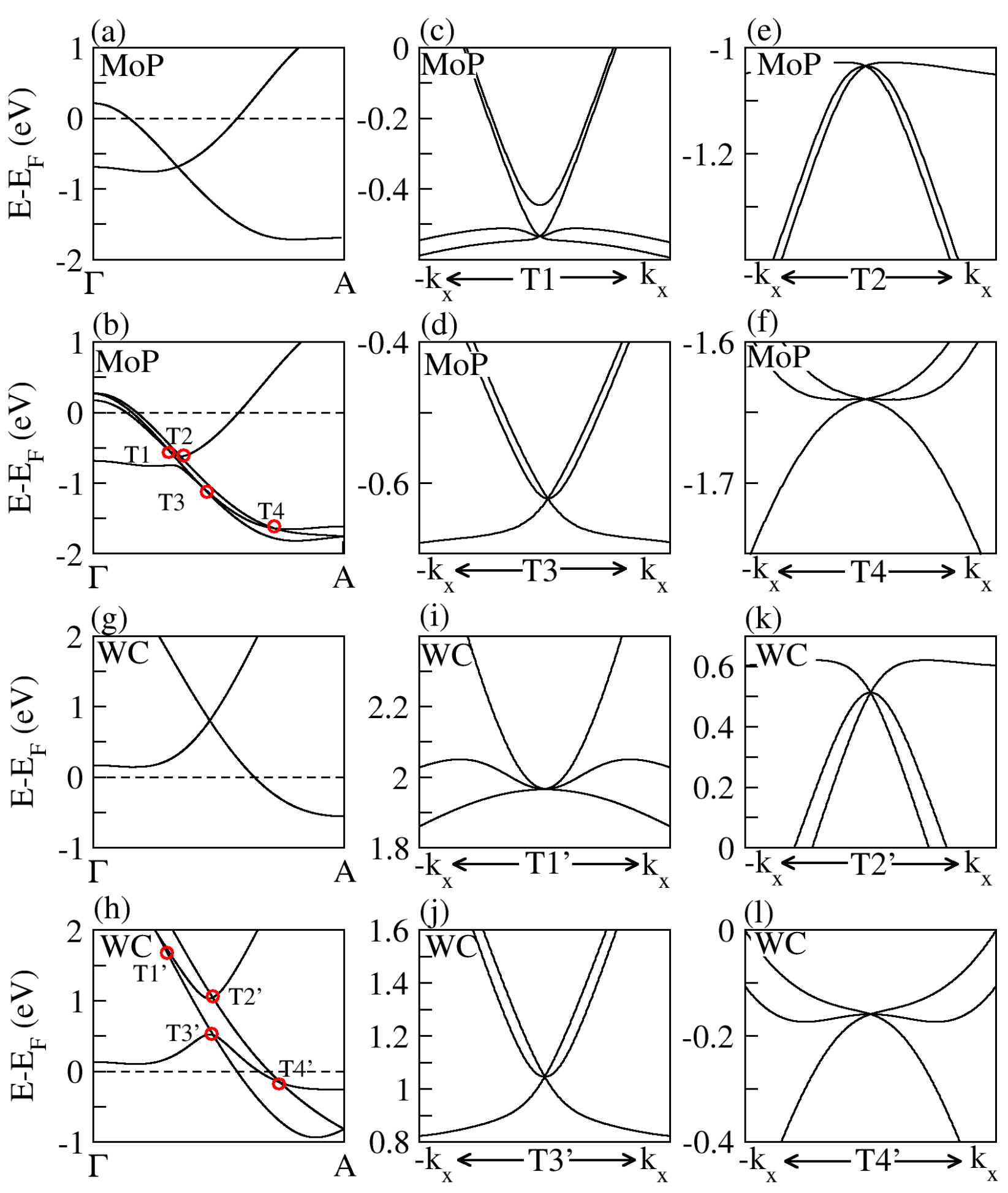

**Supplementary Figure 1 | Triple point in MoP and WC**. (**a**) and (**b**) Energy dispersion along the high symmetry line Γ-A without and with SOC, respectively. The four triple points are highlighted by the red circles. (**c-f**) Local energy dispersion with crossing of the four triple points along the $k_x$ direction. (**g-l**) Similar plots for WC.

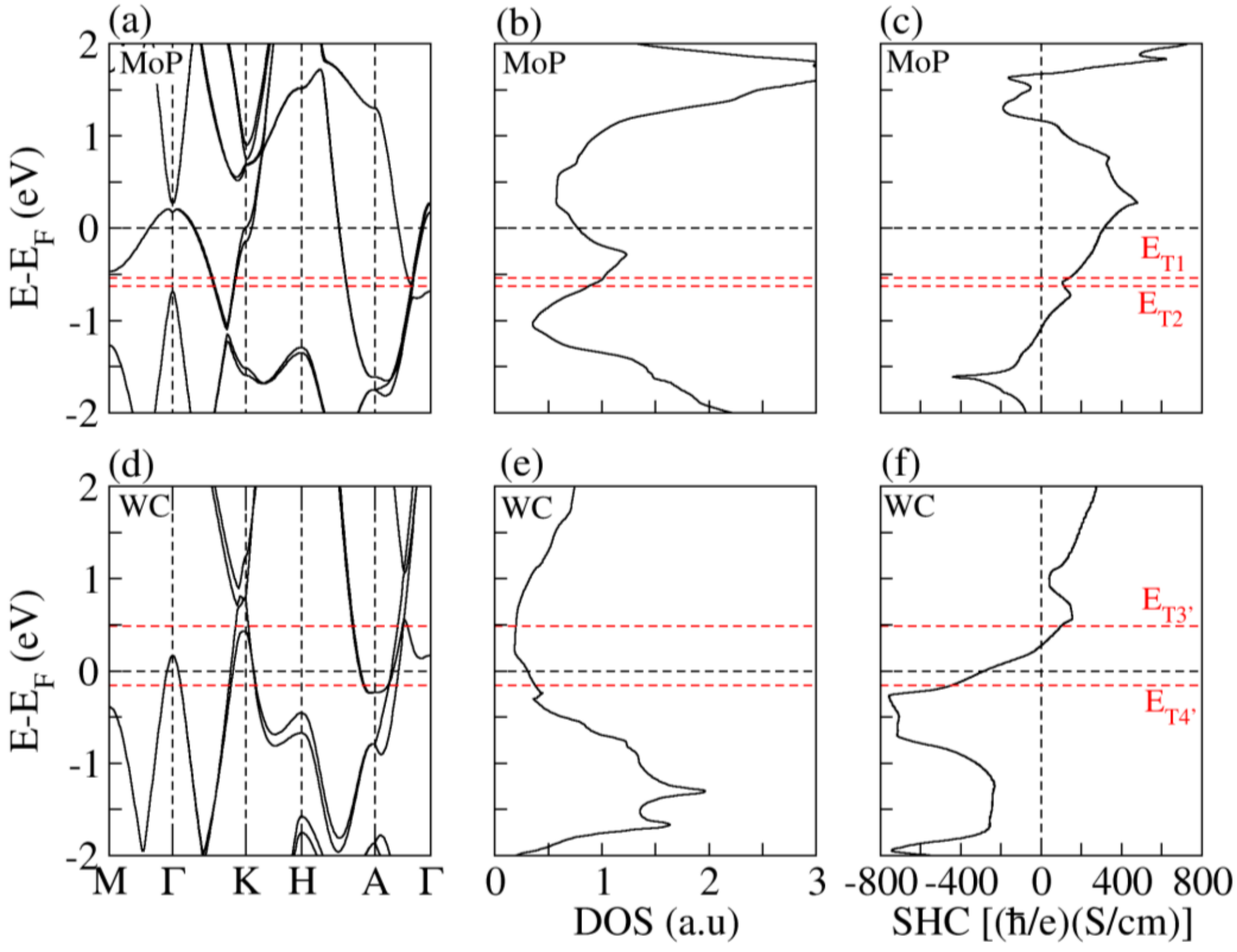

**Supplementary Figure 2 |** Energy dispersion along high symmetry lines, density of states, and energy dependent spin Hall conductivity (SHC) in (**a-c**) for MoP and (**d-f**) for WC.

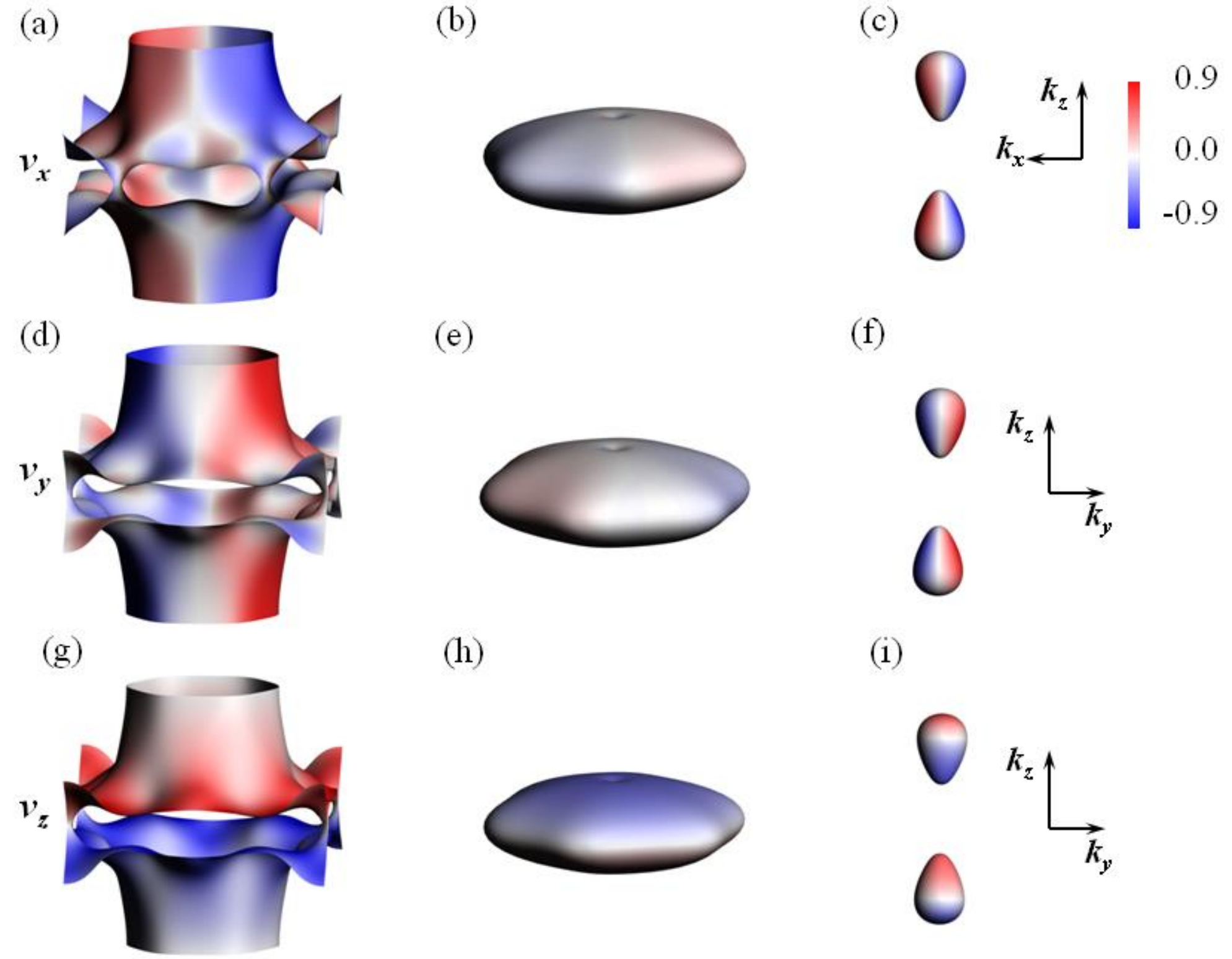

**Supplementary Figure 3 | Contributions of the Fermi velocity, $\mathbf{v}_F$, from the charge carriers from different parts of the Fermi surfaces**. The three components of (**a-c**) $\mathbf{v}_x$, (**d-f**) $\mathbf{v}_y$, and (**g-i**) $\mathbf{v}_z$ are given separately in the top, middle, and bottom panels, respectively. The maximum Fermi velocity can reach up to $0.9 \times 10^6$ m/s. The split Fermi velocity on each pair of FS by SOC is almost the same, we just show one branch for each type of FSs. The color bar is in the unit of $10^6$ m/s

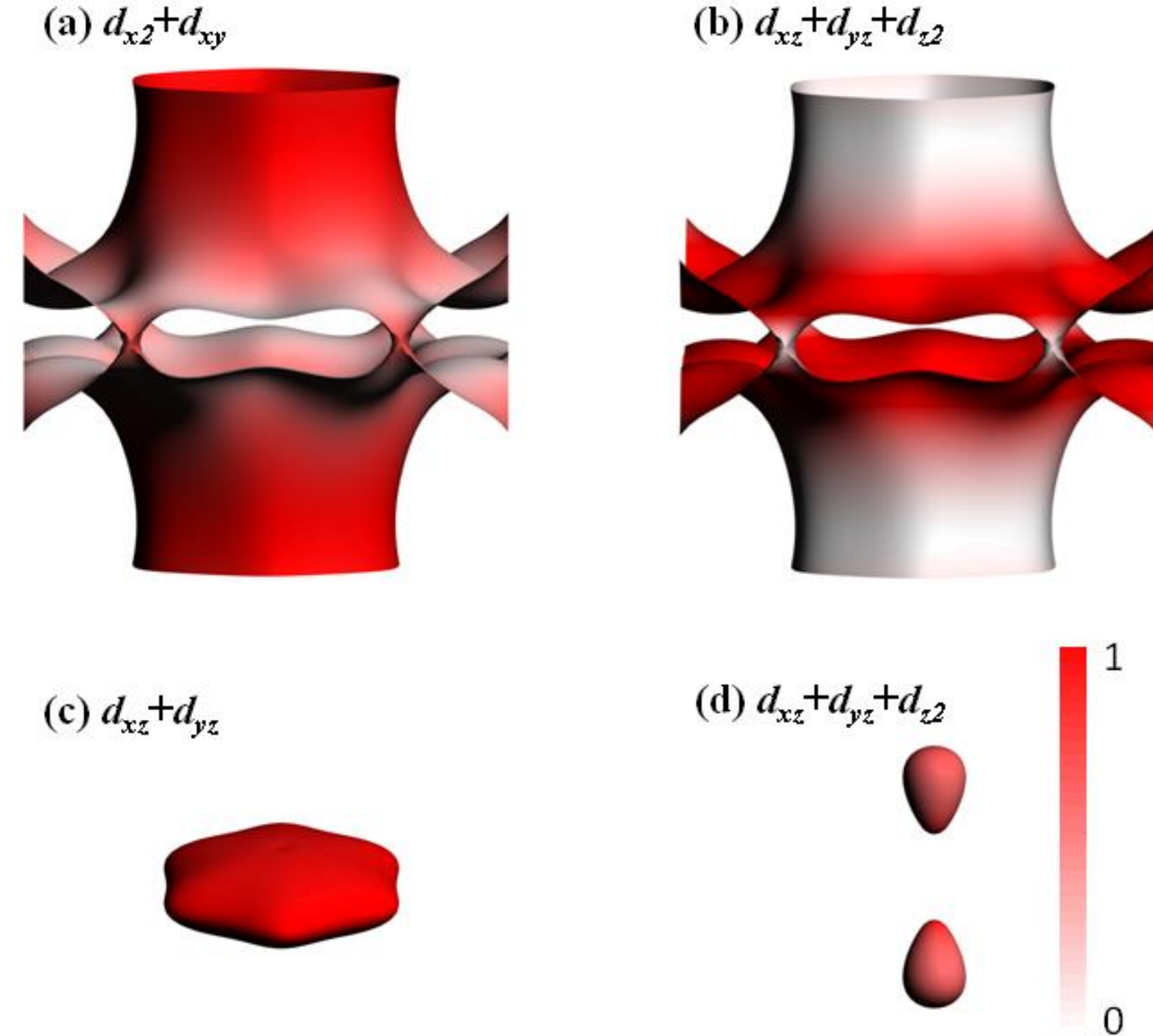

**Supplementary Figure 4 | Orbital distribution on the Fermi surface**. (**a**) Near the top and bottom of the BZ, the open FSs have mainly from Mo-$d_{x2}$+$d_{xy}$ character. (**b**) Around the center of the BZ, the open FS are dominated by Mo-$d_{xz}$+$d_{yz}$+$d_{xz}$. (**c**) Hole and (**d**) electron pockets are dominated by Mo-$d_{xz}$+$d_{yz}$ and Mo-$d_{xz}$+$d_{yz}$+$d_{z2}$, respectively. Spin splitting negligibly affects the orbital distribution of the FSs.

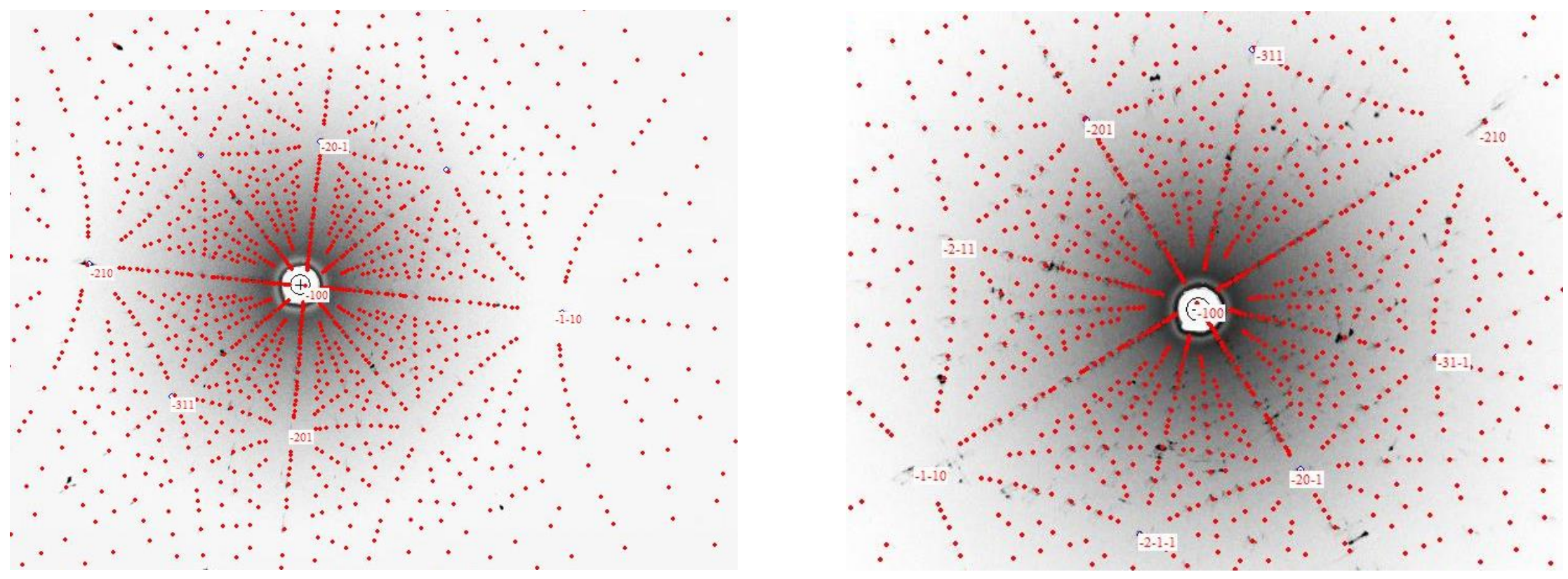

**Supplementary Figure 5 | Representative Laue diffraction patterns of as-grown (100) facets of the MoP single crystals S1 (left) and V1 (right).** The red spots and assigned Miller indices show the calculated diffraction pattern of the $P\bar{6}m2$ hexagonal-space group. After identifying the direction, the crystals were cut along desired crystallographic axes for further measurements.

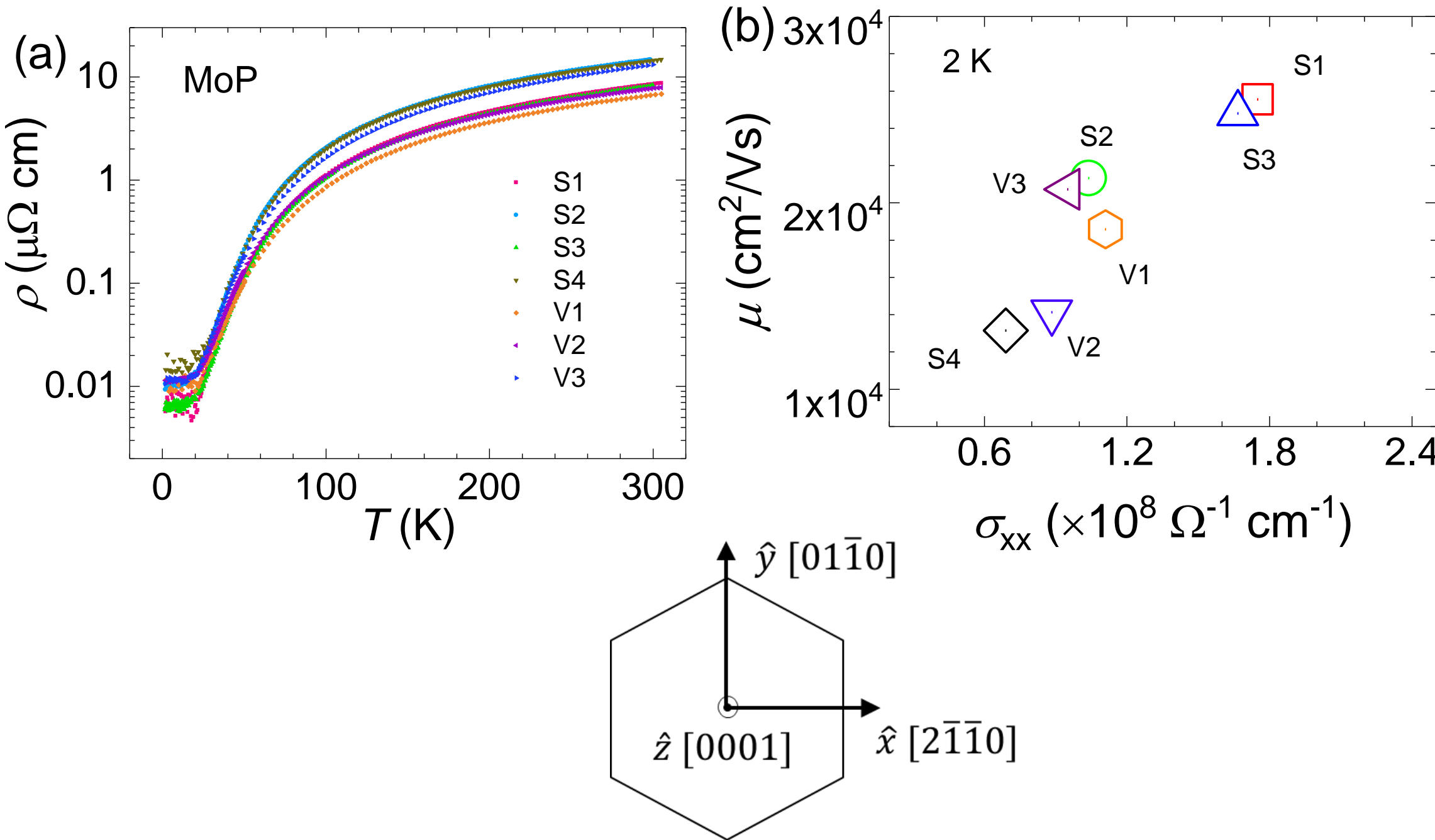

**Supplementary Figure 6 |** (**a**) Temperature-dependent resistivity (left panel), $\rho$, of seven crystals synthesized in two batches (namely S and V). (**b**) Residual conductivity, $\sigma_{xx} = 1/\rho_{xx}$ at $B = 0$, and, mobility, $\mu$ of various MoP crystals from two different batches (S,V) at 2 K. The lower panel shows a sketch defining the $\hat{\mathbf{x}}$, $\hat{\mathbf{y}}$, and $\hat{\mathbf{z}}$ directions in terms of hexagonal planes, where $\hat{\mathbf{x}}$ is $2\bar{1}\bar{1}0$, $\hat{\mathbf{y}}$ is $01\bar{1}0$, and $\hat{\mathbf{z}}$ is $0001$.

**Supplementary Table 1 |** Summary of electrical transport parameters of the seven different crystals investigated.

| Crystal | Dimension (error: ± 0.005 mm) [length × width × height ($l$×$w$×$h$) mm$^3$] | Measurement Geometry | $\rho$ at 2K (nΩ cm) | *RRR* | $\mu$ at 2K ($10^4$ cm$^2$V$^{-1}$s$^{-1}$) | $n$ at 2K ($10^{22}$ cm$^{-3}$) |
|---|---|---|---|---|---|---|
| S1 | 1.0×0.28×0.20 | $\mathbf{I}\|\|\hat{\mathbf{x}}$, $\mathbf{B}\|\|\hat{\mathbf{z}}$ | 5.7 | 1526 | 2.6 | 4.3 |
| S2 | 1.0×0.25×0.23 | $\mathbf{I}\|\|\hat{\mathbf{z}}$, $\mathbf{B}\|\|\hat{\mathbf{x}}$ | 9.6 | 1520 | 2.1 | 3.1 |
| S3 | 0.9×0.23×0.16 | $\mathbf{I}\|\|\hat{\mathbf{x}}$, $\mathbf{B}\|\|\hat{\mathbf{z}}$ | 6 | 1370 | 2.5 | 4.3 |
| S4 | 1.0×0.1×0.05 | $\mathbf{I}\|\|\hat{\mathbf{x}}$, $\mathbf{B}\|\|\hat{\mathbf{z}}$ | 14 | 1013 | 1.3 | -- |
| V1 | 0.7×0.22×0.09 | $\mathbf{I}\|\|\hat{\mathbf{z}}$, $\mathbf{B}\|\|\hat{\mathbf{y}}$ | 9.2 | 740 | 1.9 | 3.7 |
| V2 | 1.0×0.25×0.08 | $\mathbf{I}\|\|\hat{\mathbf{x}}$, $B\|\|\hat{\mathbf{z}}$ | 11 | 700 | 1.4 | 3.9 |
| V3 | 0.9×0.15×0.12 | $\mathbf{I}\|\|\hat{\mathbf{x}}$, $\mathbf{B}\|\|\hat{\mathbf{y}}$ | 11 | 1240 | 2.1 | 2.9 |

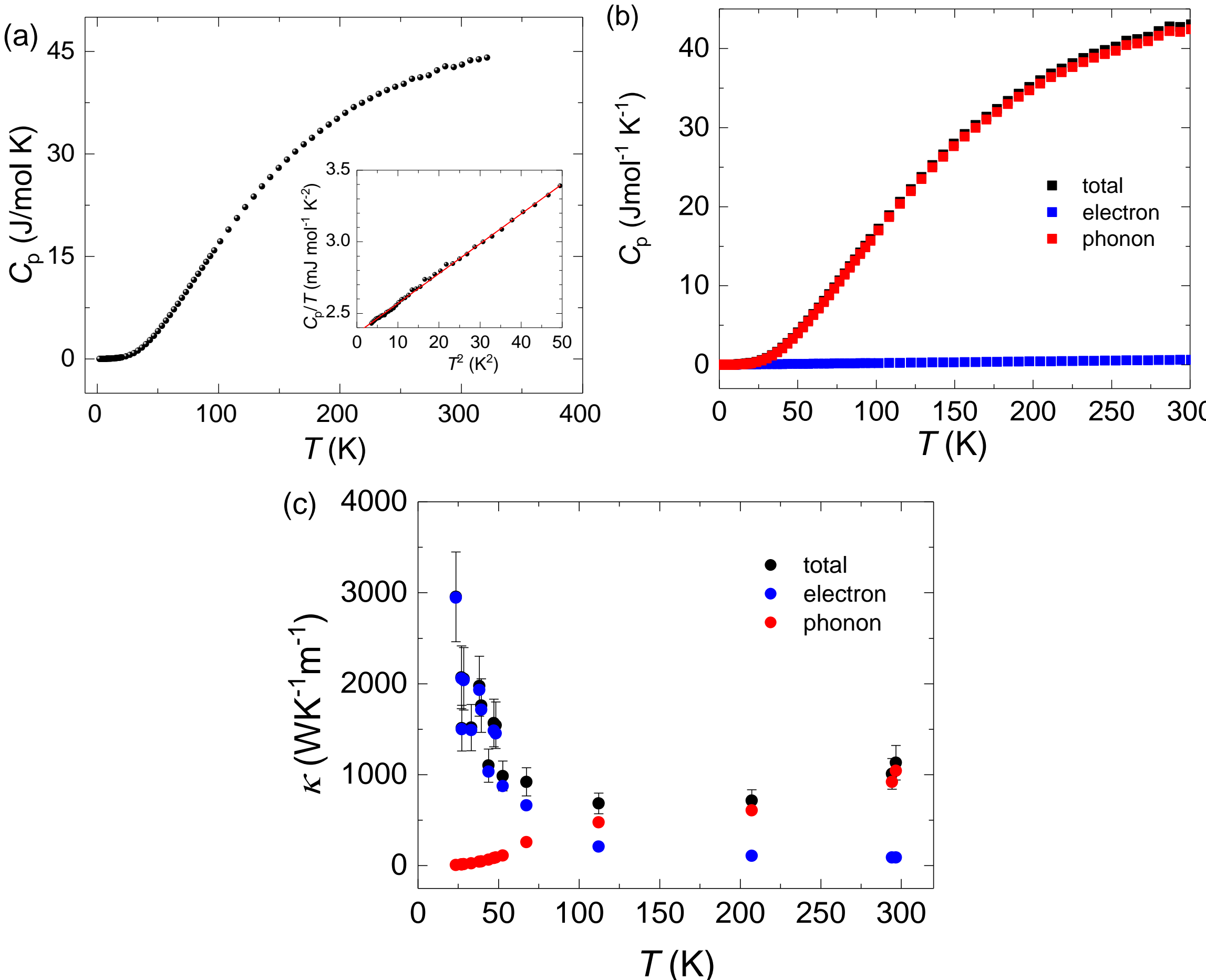

**Supplementary Figure 7 |** (**a**) Temperature dependence of the specific heat $C_p(T)$ of MoP. The inset shows a fit (red line) to the data using $C_p(T)/T = \gamma + \beta T^2$. (**b**) Measured temperature dependent total heat capacity, $c_p$ (total), separated phononic part, $C_{ph}$, and electronic part, $C_{el}$. (**c**) Temperature dependent measured total thermal conductivity, κ (total), calculated phononic (phonon) part from heat capacity and the subtracted electronic part from the measured data. The error bar in the total thermal conductivity data is the standard deviation of systematic error estimated from the uncertainty in thermal gradient and heater voltages.

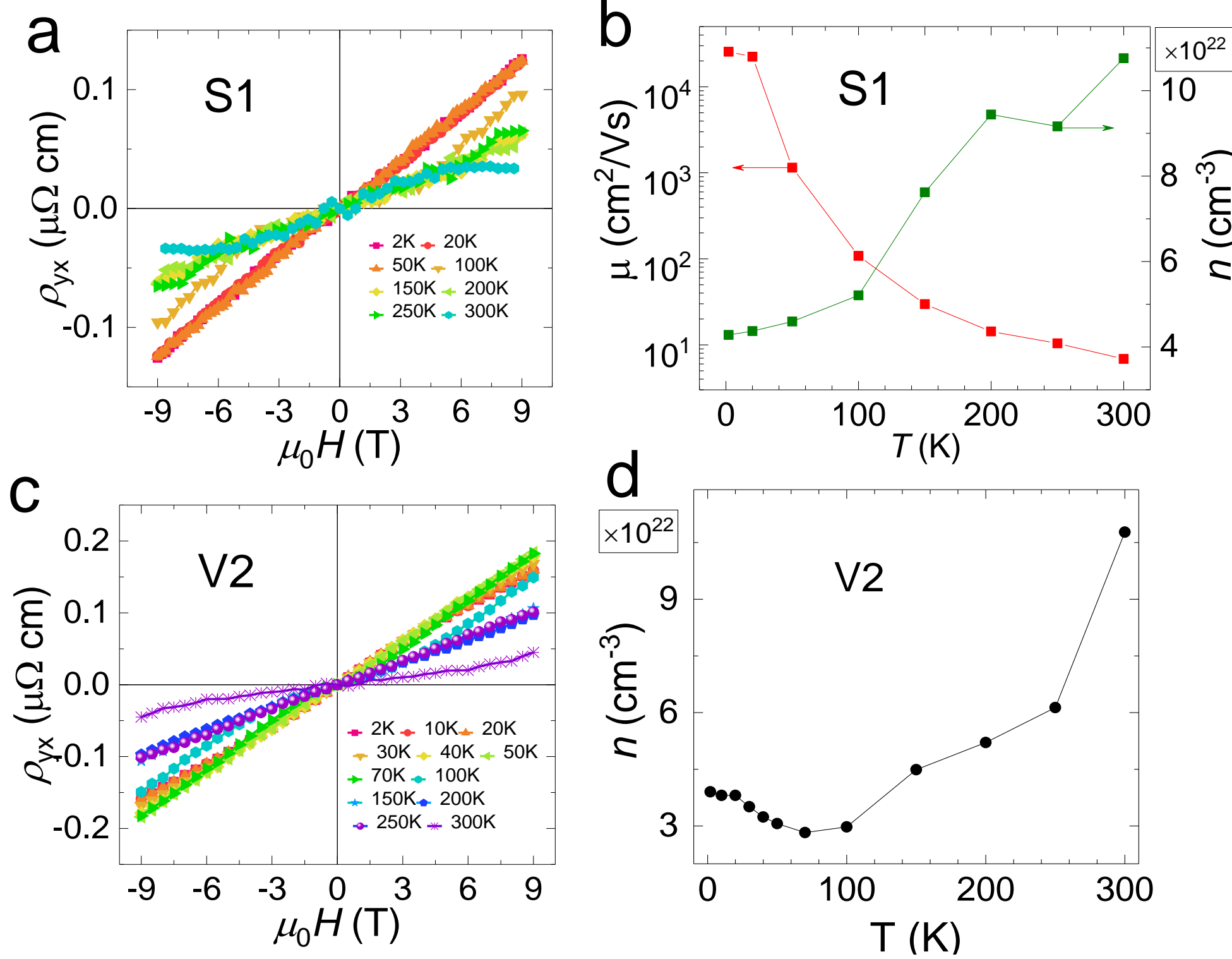

**Supplementary Figure 8 |** (**a**) and (**c**) Hall resistivity, $\rho_{yx}$, at various temperatures for crystals S1 and V2. (**b**) and (**d**) estimated temperature dependent mobility, $\mu$, and charge carrier density, $n$, extracted from the corresponding Hall data using the relations for the Hall coefficient, $R_H = \rho_{yx}/B$, $\mu = R_H/\rho_{xx}$ and $n = 1/eR_H$.

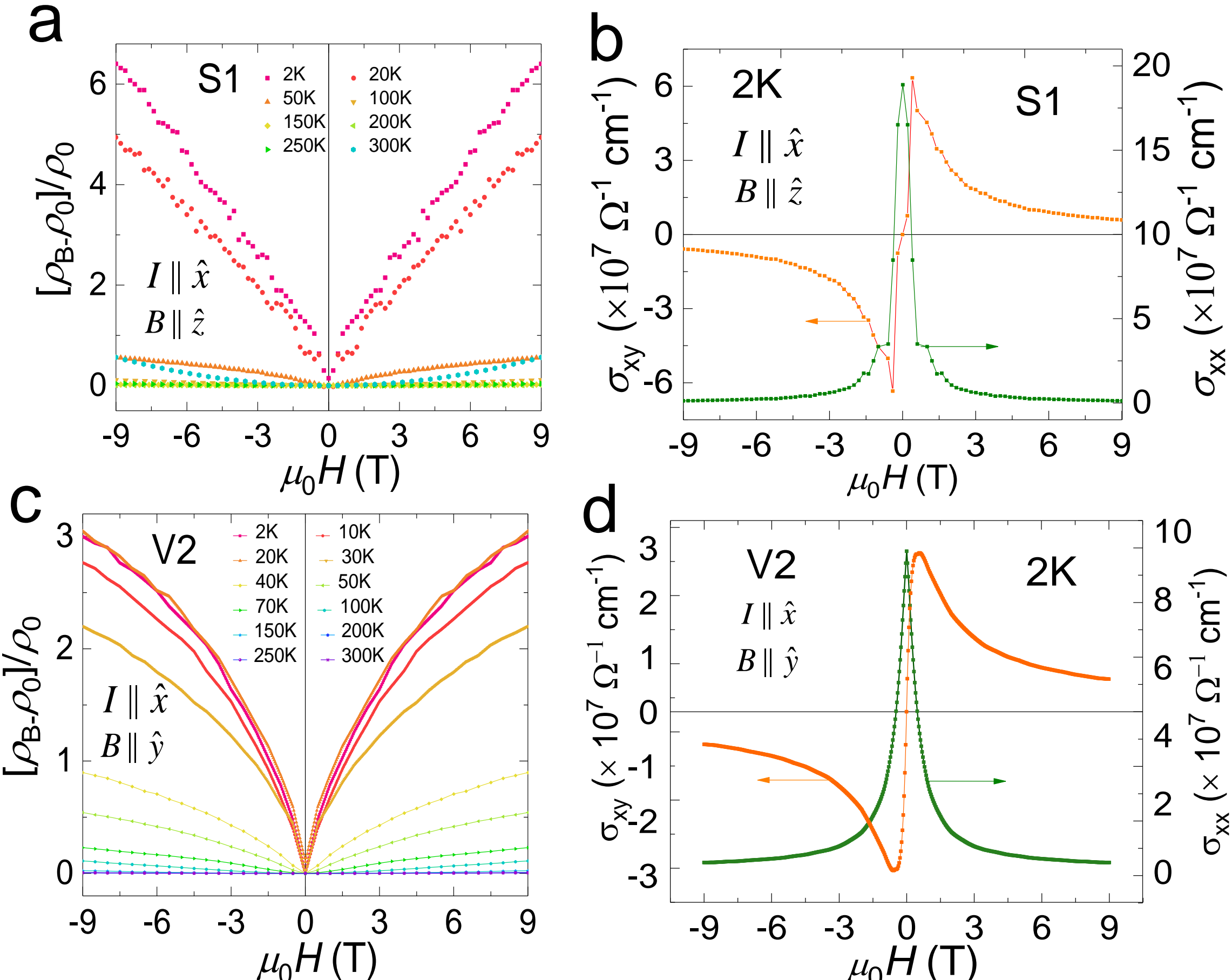

**Supplementary Figure 9 |** (**a**) and (**c**) Field dependence of the transverse magnetoresistance of crystals S1 and V2. (**b**) and (**d**) corresponding calculated Hall conductivity, $\sigma_{xy}$, and transverse conductivity, $\sigma_{xx}$, extracted from the relations $\sigma_{xy} = \rho_{yx}/(\rho_{xx}^2 + \rho_{yx}^2)$ and $\sigma_{xx} = \rho_{xx}/(\rho_{xx}^2 + \rho_{yx}^2)$. At low temperature, the magnetoresistance steeply increases at low fields hinting at anti-localization of charge carriers. This effect generally appears in compounds, which possess strong spin-orbit coupling[1,2].

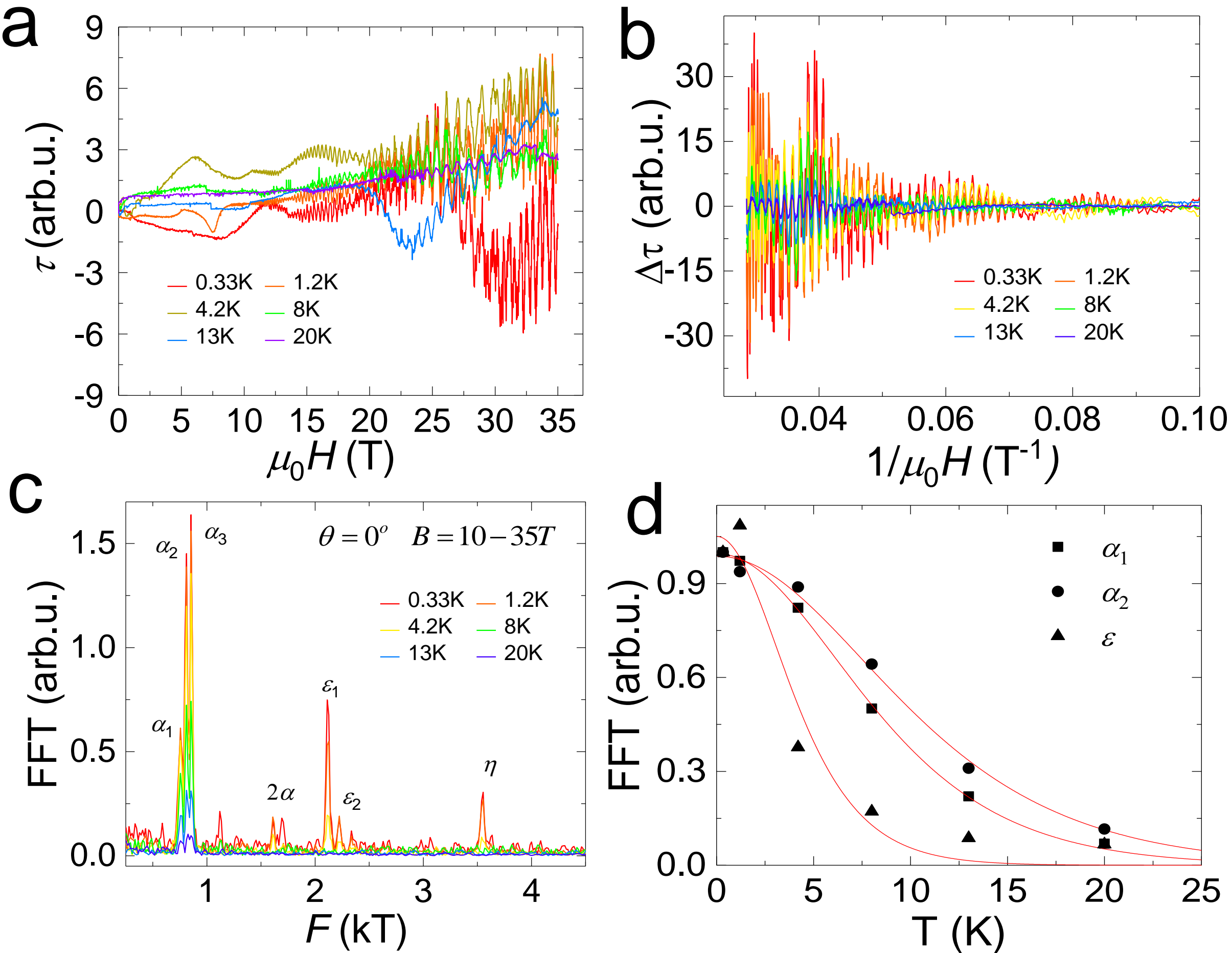

**Supplementary Figure 10 |** (**a**) Magnetic torque data at different temperatures. (**b**) dHvA oscillations between 10 and 35 T after subtraction of a smooth background at each temperature. These oscillations are periodic in $1/B$, where $B = 1/\mu_0 H$. (**c**) Fast Fourier transforms (FFTs) of the data shown in (b), which reveals six main frequencies $\alpha_1$, $\alpha_2$, $\alpha_3$, $\varepsilon_1$, $\varepsilon_2$, and $\eta$. (**d**) Temperature dependent normalized FFT amplitude of the dHvA signals corresponding to $\alpha_1$, $\alpha_2$, and $\varepsilon_2$, with fits using the Lifshitz-Kosevich formula for extracting the effective masses.

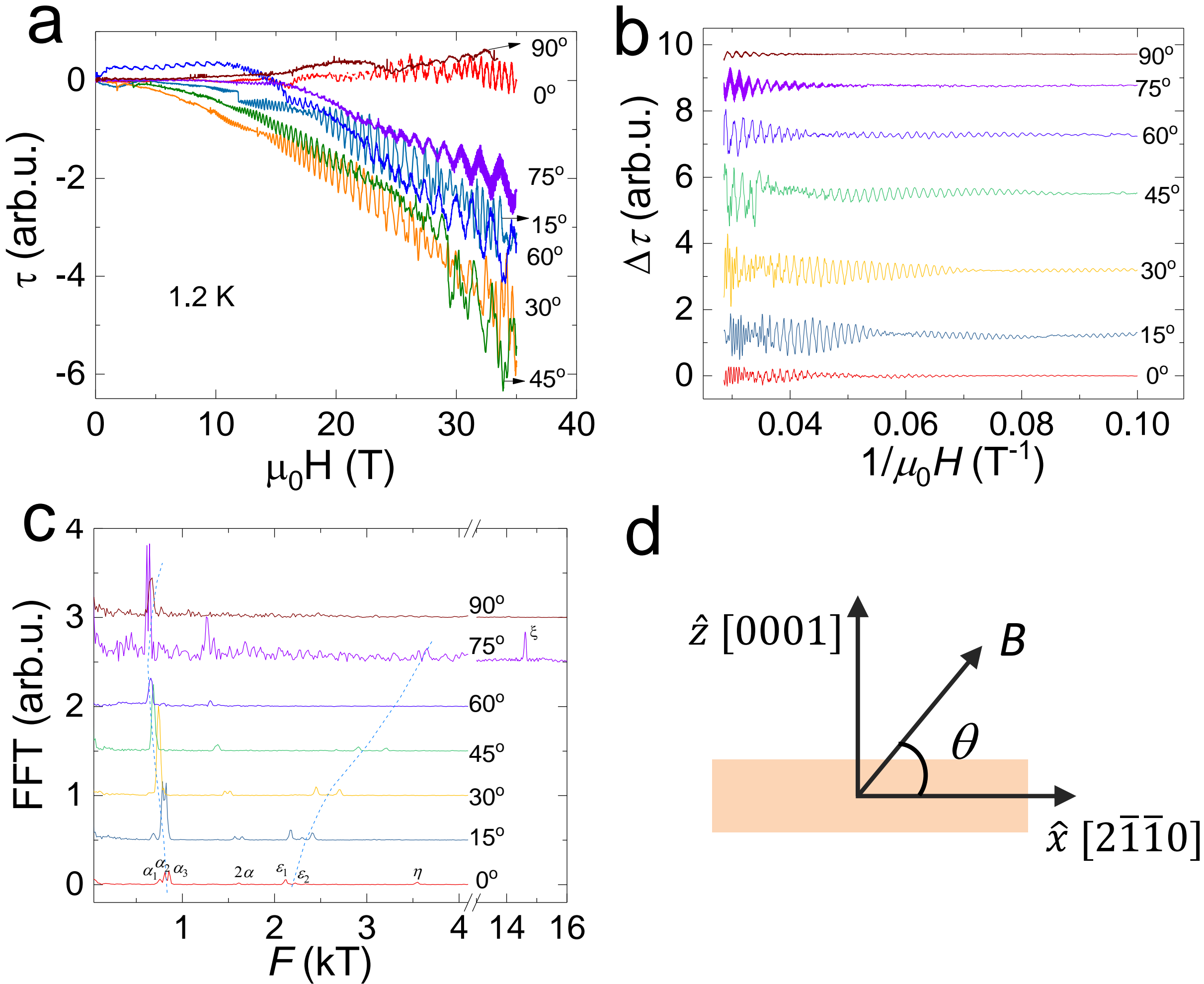

**Supplementary Figure 11 |** (**a**) Angular dependent magnetic torque data at $T$ = 1.2 K. (**b**) dHvA oscillations between 10 and 35 T after subtraction of a smooth background. (**c**) Fast Fourier transforms (FFTs) of the data shown in **b** in which dotted lines guide the shift of the frequencies with angle. At $\theta = 70^{\circ}$, a very high frequency appears at 14560 T which is denoted by $\xi$. (**d**) Schematic of the field rotation direction with respect to the crystallographic direction.

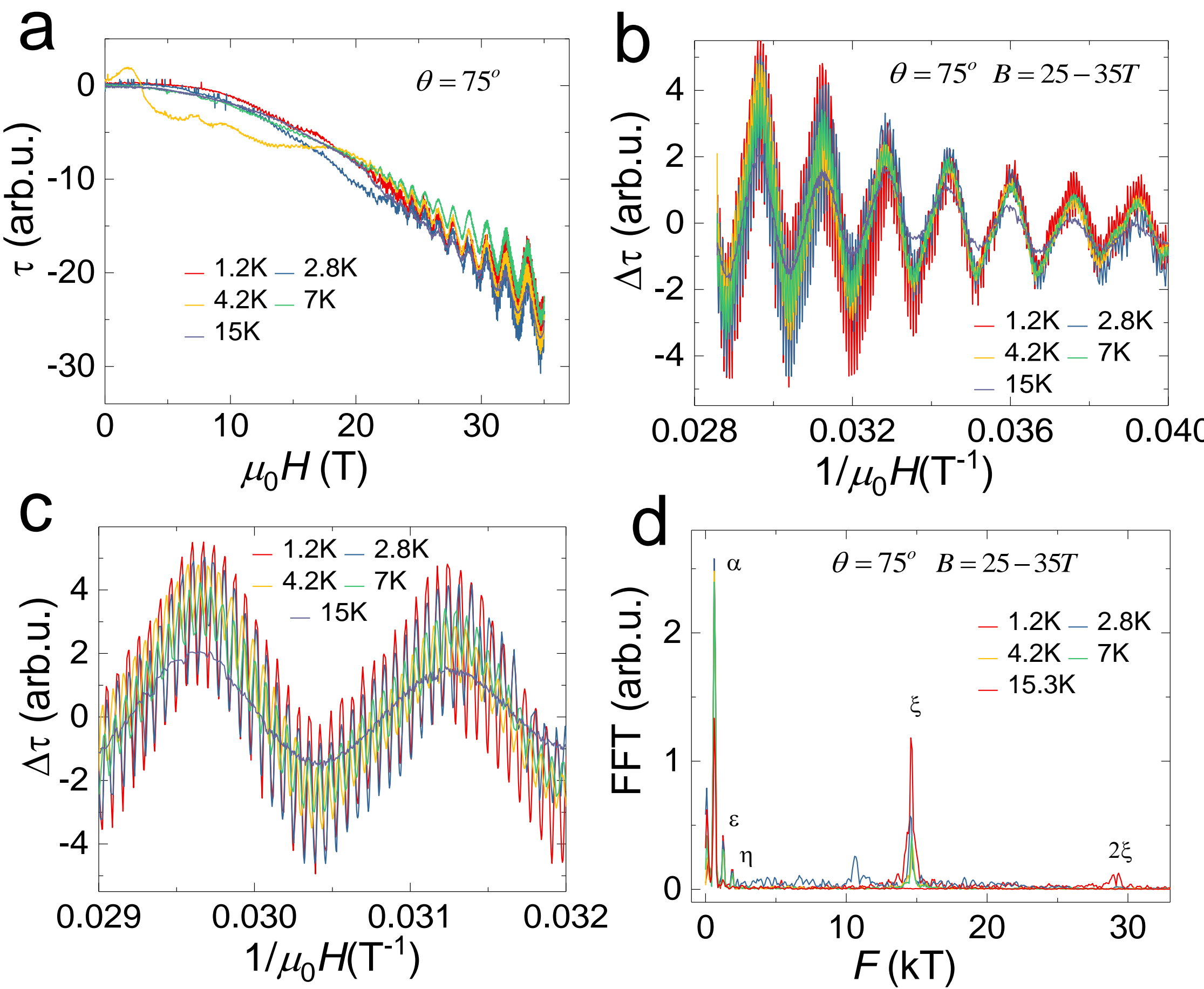

**Supplementary Figure 12 |** (**a**) Temperature dependent magnetic torque data at $\theta = 75^o$. (**b**) dHvA oscillations between 25 and 35 T after subtraction of a smooth background. (**c**) Zoom of the oscillations shown in (b) in which the high frequency oscillations are clearly visible below 15 K. (**d**) FFT of the data shown in **b**. Besides the frequencies shown in Supplementary Fig. 10, the very high frequency appearing at 14560 T is denoted by $\xi$.

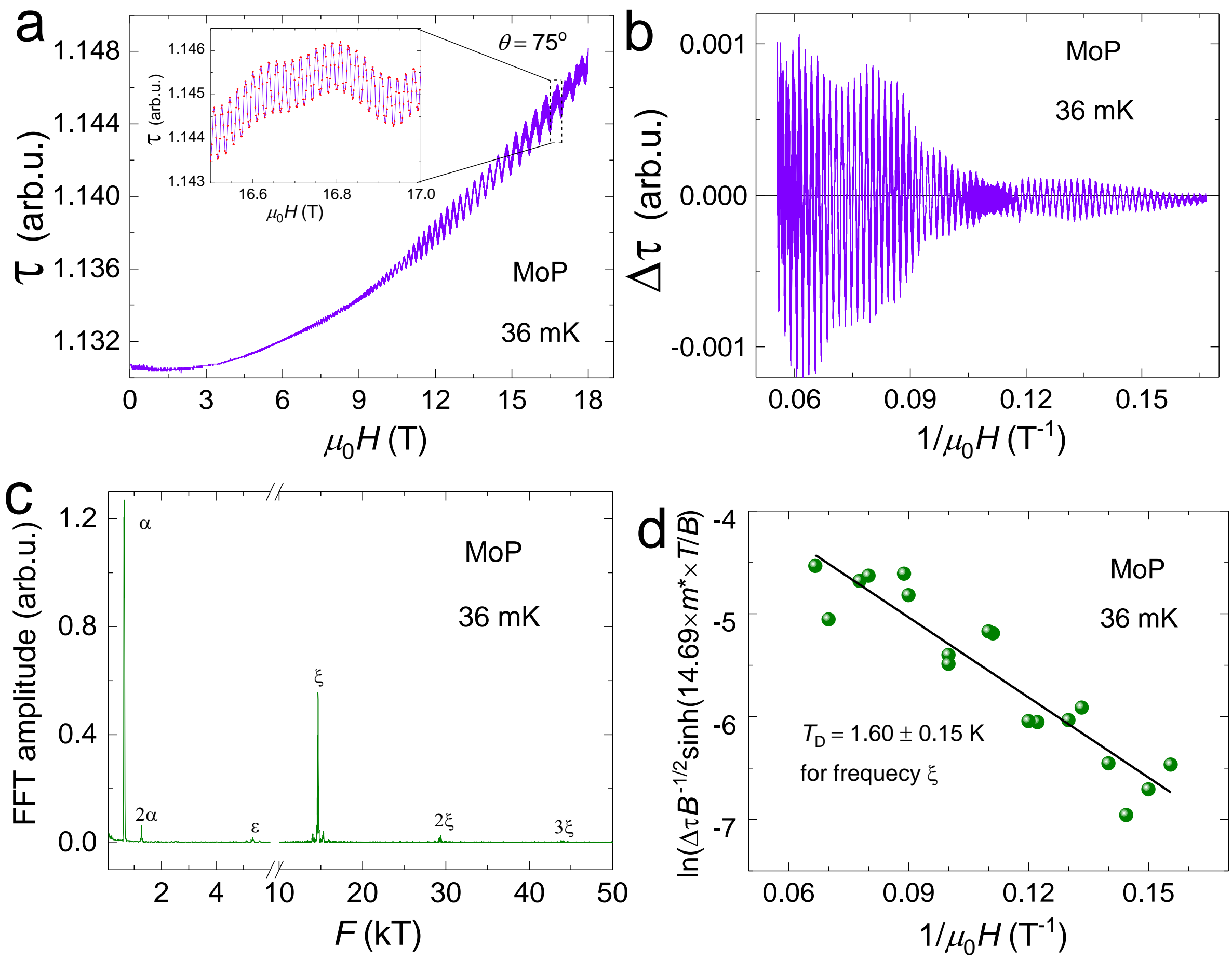

**Supplementary Figure 13 | de Haas-van Alphen (dHvA) effect of MoP obtained using an 18 T superconducting magnet at 75° on the same piece of crystal**. (**a**) Magnetic torque measurement showing dHvA oscillations at 36 mK. The inset shows an enlargement of the data between 16.5 and 17 T. (**b**) Background subtracted dHvA oscillations between 6 and 18 T. (**c**) Fourier transformation of the data in **b**. Dingle plot of the FFT amplitude of frequency ξ by choosing overlapping field windows with constant width in $1/\mu_0 H$.

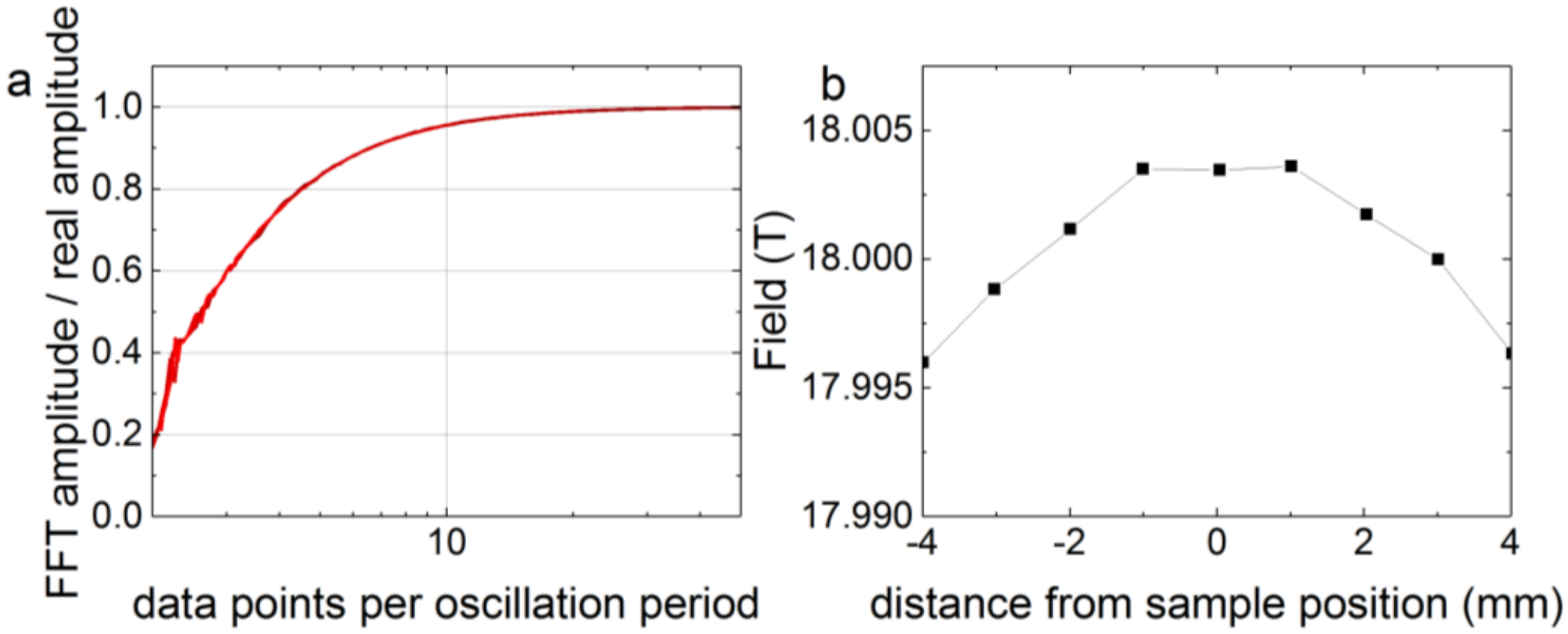

**Supplementary Figure 14 |** (**a**) Simulated damping due to limited sampling rate. (**b**) Field profile of the used 18 T superconducting magnet.

**Supplementary Table 2 |** Oscillation frequency, $F$, effective mass, $m^*$, extremal enclosed area, $A_F$, Fermi wave vector, $k_F$, Fermi velocity, $v_F$, Drude scattering time, $\tau_c$, and corresponding scattering length, $l$. These parameters are determined from the magnetic torque oscillations for the crystal V1.

| | $F$ (T) | $m^*/m_0$ | $A_F$ (Å$^{-2}$) | $k_F$ (Å$^{-1}$) | $v_F$ (m/s) ($\times 10^5$) | $\tau_c$ (s) ($\times 10^{-12}$) | $l$ (μm) |
|---|---|---|---|---|---|---|---|
| $\alpha_1$ | 750 | 0.23±0.008 | 0.0715 | 0.15 | 7.6 | 2.4 | 2.5 |
| $\alpha_2$ | 810 | 0.28±0.004 | 0.0773 | 0.157 | 6.5 | 3.0 | 2.6 |
| $\alpha_3$ | 855 | 0.297±0.01 | 0.0816 | 0.161 | 6.4 | 3.1 | 2.7 |
| $\varepsilon_1$ | 2120 | 0.83±0.09 | 0.202 | 0.254 | 3.5 | 9.0 | 4.2 |
| $\varepsilon_2$ | 2220 | 0.56±0.027 | 0.21 | 0.26 | 5.4 | 8.9 | 4.3 |
| $\eta$ | 3550 | 0.73±0.05 | 0.339 | 0.33 | 5.2 | 7.9 | 5.4 |
| $\xi$ | 14560 | 1.12±0.09 | 1.39 | 0.67 | 6.9 | 12 | 11 |

## Supplementary Note 1

**Electronic band structure and triple point fermion.** The electronic band structure along high symmetry lines of MoP is given in Supplementary Fig. 1b in the main text where we can see that the valence band and conduction bands cross the Fermi energy ($E_F$). At $E_F$, the character of Mo $d$ orbitals dominate over the $p$ orbitals of P while they are strongly hybridized. In the absence of spin orbit coupling (SOC), the band inversion between Mo-$d_{z2}$ and -$E_g$ orbitals forms a three-fold degeneracy on the line of $\Gamma$-A (Supplementary Fig. 1a), which is protected by $C_3$ rotation symmetry. Due to the lack of the inversion symmetry in MoP, all such bands are spin non-degenerate at generic $k$ points when including SOC. The bands along $\Gamma$-$A$ are reconstructed into two classes by $M_z$ mirror symmetry, two doubly degenerated $|J_z| = 1/2$ bands and two non-degenerated $|J_z| = 3/2$ bands. The crossing between these two types of bands leads to four triple points (TPs) (Supplementary Fig. 1b) which are protected by $C_3$ symmetry[3,4]. Further, the dispersions of the four TPs along the $k_x$ direction are shown in Supplementary Fig. 3c-f, where one can see that all these TPs are formed by quadratic bands and further splitting give rise to Weyl points. For comparison, we also analyzed the triple points in WC. Similar to MoP, there are also four triple points along the Γ-A direction. Due to one less valence electron in WC, the three triple points are above the $E_F$ and one below $E_F$, see Supplementary Fig. 1g-l. As compared to MoP, one triple point is very close to $E_F$, which is only 0.15 eV below $E_F$, as shown in the Supplementary Fig. 1k-l.

## Supplementary Note 2

**Spin Hall conductivity.** To check the influence of the triple points, we calculated the spin Hall conductivity (SHC) for MoP and WC, as shown in Supplementary Fig. 2. Remarkably, we noted that the energy dependent peak position of the density of states and SHC (red dotted lines in Supplementary Fig. 2b-c and 2e-f do not appear at the same energy in both MoP and WC. Therefore, the SHC is mainly determined by the band structure. Since the triple points appear normally along with strong entanglement between different bands in an energy window, the peak of the SHC is not lying at the triple points, but it appears near the triple points. Particularly for MoP, the spin Hall conductivity increases from the triple point to the Fermi energy and has a peak slightly above $E_F$, as shown in Supplementary Fig. 2c. Thus, one can speculate that the presence of topological band may have a nontrivial influence on the transport properties of MoP.

## Supplementary Note 3

**Fermi velocity and orbital distribution on the Fermi surface.** The three components of the Fermi velocity, $\mathbf{v}_x$, $\mathbf{v}_y$, $\mathbf{v}_z$, on the Fermi surface (FS) were calculated by the approximation $v_{f,i} = \frac{\partial E}{\hbar \partial k_i}$, with $i = x$, $y$, and $z$. From Supplementary Fig. 3 one can see that the Fermi velocity is locked with the lattice momentum. The maximum Fermi velocity appears on the two open FSs, and parts of them are larger than $0.9 \times 10^6$ $\mathrm{ms^{-1}}$ that leads to the high mobility in the transport (Supplementary Fig. 3). Around the center of the BZ, the Fermi velocity is mostly dominated by $\mathbf{v}_z$ while $\mathbf{v}_x$+$\mathbf{v}_y$ dominate near the top and bottom of the BZ. The hole pockets centered at the Γ point contribute only to $\mathbf{v}_z$, and the other two components are almost zero. Even though the electron pockets also have a tiny charge carrier concentration, they contribute appreciably to the Fermi velocity. The open FSs and hole pockets mainly dominate the z components of the Fermi velocity, while the $\mathbf{v}_x$ and $\mathbf{v}_y$ are dominated by the open FSs.

In term of orbitals, the FSs with large $v_x$ and $v_y$ are mainly influenced by the Mo-$E_g$+$d_{xy}$ orbitals, while $\mathbf{v}_z$ is dominated by the Mo-$d_{xz}$+$d_{yz}$ orbitals (Supplementary Fig. 4). The relation between the Fermi velocity and orbital distribution of the FSs simply implies that the interlayer hopping of charge carriers between Mo and P determines $\mathbf{v}_z$, while intralayer overlap between Mo controls $\mathbf{v}_x$ and $\mathbf{v}_y$.

## Supplementary Note 4

**Determination of crystallographic directions.** The quality of different MoP crystals and the corresponding orientations were analyzed by white beam backscattering Laue X-ray diffraction at room temperature. The samples show very sharp spots, which can be indexed by a single pattern, suggesting excellent quality of the grown crystals without any twining or domains. Supplementary Fig. 5 represents Laue patterns of crystals superposed with simulated patterns. The data fit with hexagonal $P\bar{6}m2$ crystal structure with lattice parameters $a = b = 3.22$ Å and $c = 3.19$ Å.

## Supplementary Note 5

**Electrical resistivity.** After characterization of two batches of MoP crystals, seven crystals were cut in similar bar shapes but in desired crystallographic directions (see Supplementary Table 1) and four contacts for resistivity and five contacts for Hall resistivity were made using 25 μm Pt wire with Ag paint. Supplementary Table 1 summarizes the crystals' dimension, direction of current, $I$, magnetic field, $B$, residual resistivity at $T$ = 2 K, residual resistivity ratio ($\rho_{300K}/\rho_{3K}$), mobility, $\mu$, and charge carrier density, $n$, for each crystal. We can see that $\rho$ is independent from the crystallographic direction showing isotropic transport in MoP.

## Supplementary Note 6

**Specific heat.** The specific heat, $C_p$, was measured on six selected large crystals (Supplementary Fig. 7a). The Dulong-Petit limit $3nR$ ($n$ = number of atoms, $R$ = molar gas constant) is not yet reached at 320 K, indicating relatively strong chemical bonding in MoP. No anomalies from phase transitions are observed in the range covered. The inset shows a $C_p/T$ vs. $T^2$ representation of the data below $T$ = 7 K. The line indicates a fit of the data to $C_p(T) = \gamma T + \beta T^3$ (inset of the Supplementary Fig. 7), with the conduction electron specific heat $\gamma T$ and the phonon specific heat in the Debye $T^3$ approximation. For the (Sommerfeld) coefficient $\gamma$, we obtain 2.37 mJ mol$^{-1}$ K$^{-2}$ and $\beta$ corresponds to a Debye temperature $\Theta_D$ = 573 K. The magnetic field dependence of $C_p(T)$ measured at $\mu_0H$ = 9 T is insignificant. The theoretical integrated density of states (DOS) at $E_F$ is 0.78 states eV$^{-1}$ f.u.$^{-1}$ which is almost equal to the experimentally determined value. A comparison with the measured linear specific heat coefficient, $\gamma$ = 2.37 mJ mol$^{-1}$ K$^{-2}$, results in an effective mass ratio $m^*/m_e$ = 1.29. Very similar values of $m^*/m_e$ are observed for simple metals such as Na,

K, Rb, Mg, Al and Cu (Ref [5]). Temperature dependent total thermal conductivity, κ (total), calculated phononic (phonon) part using the specific-heat data and remaining electronic part.

The total measured heat capacity, $C_{total}(T) = C_{el}(T) + C_{ph}(T)$, contains an electronic part, $C_{el}(T)$, and a phonon part, $C_{ph}(T)$. At low temperatures ($T < 10$ K), $C_{el}(T)$ and $C_{ph}(T)$ can be separated by a fit using $C_{el}(T) = \gamma T$ and $C_{ph}(T) = \beta T^3$ ($T < 10$ K).

In contrast to $C_{ph}(T) = \beta T^3$, $C_{el}(T) = \gamma T$ is valid over the full temperature range.
Therefore, the phonon contribution is given by
$C_{ph}(T) = C_{total}(T) - \gamma T$, as shown in Supplementary Fig. 7b.
The total measured thermal conductivity, $\kappa_{total}$, is attributed to the electronic part, $\kappa_{el}$, and the phononic part, $\kappa_{ph}$, i.e.,

$$\kappa_{total}(T) = \kappa_{el}(T) + \kappa_{ph}(T) \qquad (1)$$

Assuming that the Wiedemann-Franz law holds at 300 K, the electronic contribution can be calculated as $\kappa_{el}$ (300 K) = $L_0\sigma$(300 K) = 89 $WK^{-1}m^{-1}$, where $\sigma$ is the measured electrical conductivity at 300 K ($1.25 \times 10^7$ $\Omega^{-1}$ $m^{-1}$) and $L_0$ ($2.41\times 10^{-8}$ $W\Omega$ $K^{-2}$) is the Lorenz number. Consequently, the phononic contribution of $\kappa$ at 300 K is:

$$\kappa_{ph}(300\text{ K}) = \kappa_{total}(300\text{ K}) - \kappa_{el}(300\text{ K}) = 1132-89 = 1043\ WK^{-1}m^{-1}$$

From the kinetic relation, $\kappa_{ph}$ ($T$) is written as:

$$\kappa_{ph}(T) = 1/3\ C_{ph}(T)\ \upsilon^2\ l_{ph}(T), \qquad (2)$$

where $\upsilon$ is the sound velocity and $l_{ph}$ the phonon mean free path. For simplicity, we assume that $l_{ph}$ is independent of temperature. However, $l_{ph}$ is generally increasing with decreasing temperature. By this assumption, we can now calculate the temperature-independent 1/3 $\upsilon^2$ $t_{ph} = \kappa_{ph}$ (300 K) / $C_{ph}$(300 K).
Using $1/3\upsilon^2 t_{ph} C_{ph}(T)$, $\kappa_{ph}$ ($T$) can be calculated over the full temperature range and, finally, $\kappa_{el}$ ($T$) can be calculated using $\kappa_{el}(T) = \kappa_{total}(T) - \kappa_{ph}(T)$. Both, $\kappa_{el\ (T)}$ and $\kappa_{ph}$ (T) are shown in Supplementary Fig. 7c.

Now, we consider the Lorenz number $L = \kappa_{el}/\sigma T$ in MoP, which helps to characterize thermal and charge transport characteristics. In conventional conductors, the Wiedemann-Franz law holds approximately, i.e., $L$ equals the Lorenz number $L_0 = \pi^2 k_B^2/(3e^2)$. The experimentally extracted $\kappa_{el}$ as a function of $T$ is given in Supplementary Fig. 7c.

## Supplementary Note 7

**Magnetic torque**. The torque magnetization was measured using capacitive cantilevers mounted on rotatable platforms at the High Field Magnet Laboratory (HFML) in Nijmegen and at the High Magnetic Field Laboratory in Dresden (HLD). The method is based on measuring the change in the capacitance between a reference plate and a flexible CuBe cantilever, on which the sample is attached, as a function of magnetic field. This change in capacitance is due to a slight bending of the cantilever caused by the force exerted by the sample in an external magnetic field. The capacitance was measured using Andeen-Hagerling 2700A capacitance bridges, and the angle with respect to the field was measured using a Hall probe via a Stanford Research SR830 lock-in amplifier (Nijmegen) and via a calibrated stepper motor in Dresden. Initially, the field was applied perpendicular to the cantilever ($B \parallel \hat{\mathbf{x}}$ ($\theta = 0^{o}$)) and then the sample was rotated stepwise until the sample and cantilever were parallel to the applied field ($B \parallel \hat{\mathbf{z}}$ ($\theta = 90^{o}$)).

The samples were single crystalline and the perpendicular magnetization component (torque) was measured. The parallel magnetization component (when the sample is placed in a field gradient) was found to be extremely small. Due to the geometry of the experiment, the force from the sample is proportional to $Ae_0/C$, where $C$ is the measured capacitance, $A$ is defined by the dimensions and force constant of the cantilever, and $e_0$ is the vacuum permittivity. However, since only relative changes of capacitance (and hence force) are considered in this work, the data are shown on an arbitrary scale. The magnetic field was determined based on the measured current through the Bitter magnet coils, using calibration curves provided by the HFML.

## Supplementary Note 8

**Determination of the Dingle temperature.** The measured de Haas-van Alphen (dHvA) torque oscillations are well described by the Lifshitz–Kosevich formula [6]

$$\Delta\tau = B^{3/2} R_T R_D \sin\left[2\pi\left(\frac{F}{B} + \gamma\right)\right], \quad (3)$$

where $R_T = \frac{14.69\, m^*\, T/B}{\sinh(14.69\, m^*\, T/B)}$, $R_D = \exp(-14.69\, m^*\, T_D/B)$, in which $m^*$ is the effective mass in units of $m_0$, $T_D$ is the Dingle temperature, $\frac{2\pi^2 k_B m_0}{\hbar e} = 14.69$ TK$^{-1}$, $m_0$ is the bare electron mass, $\gamma$ is a phase factor, $F$ is the dHvA oscillation frequency, and $k_B$ and ħ are the Boltzmann and reduced Planck constants, respectively. $R_D$ and $R_T$ are related to the broadening of the Landau levels (LL) due to charge-carrier scattering the broadening of the Fermi-Dirac distribution with temperature, respectively. Analysis of the temperature dependent term, $R_T$, gives the effective mass, whereas the analysis of $R_D$ gives the value of $T_D$ in terms of the scattering time, $\tau_D$, by the relation $\tau_D = \hbar/(2\pi k_B T_D)$. $\tau_D$ is known as quantum scattering time which is assumed to be equivalent to the momentum conserving scattering time $\tau_{\mathrm{mc}}$.

To calculate $T_D$, dHvA oscillations were measured via capacitive torque magnetometry at ~36 mK in a superconducting magnet using a $^3$He/$^4$He dilution refrigerator applying magnetic fields up to 18 T. Initially, the crystal was mounted with $\mathbf{B} \parallel \hat{\mathbf{x}}$ ($\theta = 0^{\circ}$) and dHvA oscillations were measured changing the angle stepwise to track the frequency while rotating towards $\mathbf{B} \parallel \hat{\mathbf{z}}$ ($\theta = 90^{\circ}$). The observed frequencies match closely to those obtained using the 35 T magnet. Since the main interest of this measurement is to calculate $T_D$ for the largest Fermi surface which cuts a nearly circular extremal area normal to the crystallographic *c*-axis of the compound, i.e., *z*-axis, we measured data up to 18 T for an angle of 75° as shown in Supplementary Fig. 13a. The inset of this figure depicts an enlargement of the data between 16.5 and 17 T. The background subtracted dHvA oscillations between 6 to 18 T and their corresponding FFT are shown in Supplementary Fig. 13b and 13c, respectively.

Since the Dingle term varies exponentially with field (Eq. (1)), the so-called Dingle plot of $\ln\left[\Delta\tau \times B^{-1/2} \sinh(14.69\, m^*\, T/B)/\,(14.69\, m^*\, T)\right]/\,14.69\, m^*$ versus $1/B$ follows a straight line where the slope gives the value of $T_D$. For minimizing the beating effect, we selected overlapping constant field windows in 1/B for different average fields and determined the amplitude, $\Delta\tau$, from FFTs. The resulting Dingle plot is shown in Supplementary Fig. 13d. The linear slope divided by $14.69m^*$ yields the value of $T_D$, which is (1.60 ± 0.15) K for the largest Fermi surface. The oscillatory deviation from the straight line fit is due to the presence of beating caused by the two close spin-split $\xi$ frequencies.

**Absence of extrinsic effects in the determined Dingle temperature.** Independent of the method used to determine the value of $T_D$, the oscillations must be free from extrinsic factors such as field inhomogeneity, sweep rate of the magnetic field and magnetic impurities within the sample. From our sample characterizations, we can exclude magnetic impurities in our sample. The magnets used in this measurements are highly precise and homogeneous. The homogeneity of the used 18 T superconducting magnet over the sample is better than 0.2 mT at 18 T, while it is $10^{-4}$ for the 35 T magnet. It should be noted that the spatial field dependence follows approximately a parabolic function as shown in Supplementary Fig. 14b. In general, the field inhomogeneity, $\Delta H$ for a given frequency $F$ should be[6]

$$\Delta H \ll \frac{H^2}{2\pi F} \qquad (4)$$

In order to observe the dHvA frequency of 14600 T in a field of 12 T (average field of our analysis), $\Delta H$ should be much smaller than 1.6 mT. The field profile of our 18 T magnet was measured by use of a Hall sensor (Supplementary Fig. 14b). At full field the field inhomogeneity at the sample position, with the sample having dimensions of $1 \times 0.7 \times 0.2$ mm$^3$, is less than 0.2 mT (that means it is below the resolution of our Hall sensor). Correspondingly, at 12 T the inhomogeneity is less than 0.13 mT. This indeed is much smaller (by more than a factor of 10) than the estimated 1.6 mT.

Another factor which affects the value of $T_D$ is the sampling rate, i.e., the number of data points taken per oscillation period. By adjusting the sweep rate stepwise according to the magnetic field value (0.005 T/min at 4.5 T to 0.1 T/min at 18 T), we ensured the acquisition of at least 12 data points per oscillation period. To estimate the influence of the sweep rate on the dHvA amplitude, the artificial damping due to the limited sampling rate was calculated by generating a sine function of constant amplitude and simulating the loss of amplitude due to decreasing data-point density and averaging effects (Supplementary Fig. 14a). The data measured in the 35 T magnet were taken at a rather fast sweep rate. At the highest field, only 4-5 data points per period were recorded, which caused an extra damping. Therefore, we refrained from extracting $T_D$ from the data measured in the 35 T magnet.

We conclude that using the data obtained in out 18 T magnet the value of $T_D$ is neither affected by magnetic impurities nor field inhomogeneity nor by a too fast magnetic-field sweep rate.

## References

1 Shekhar, C. *et al.* Evidence of surface transport and weak antilocalization in a single crystal of the $Bi_2Te_2Se$ topological insulator. *Phys. Rev. B* **90**, 165140 (2014).

2 Xia, B. *et al.* Indications of surface-dominated transport in single crystalline nanoflake devices of topological insulator $Bi_{1.5}Sb_{0.5}Te_{1.8}Se_{1.2}$. *Phys. Rev. B* **87**, 085442 (2013).

3 Lv, B. Q. *et al.* Observation of three-component fermions in the topological semimetal molybdenum phosphide. *Nature* **546**, 627-631 (2017).

4 Zhu, Z., Winkler, G. W., Wu, Q., Li, J. & Soluyanov, A. A. Triple Point Topological Metals. *Phys. Rev. X* **6**, 031003 (2016).

5 Ashcroft, N. W. & Mermin, N. D. *Solid State Physics (Holt, Rinehart and Winston, New York, 1976)*. 403 (2005).

6 Shoenberg, D. *Magnetic Oscillations in Metals*. (Cambridge Univ. Press, 1984).